\documentclass[11pt]{article}
\pdfoutput=1

\usepackage{epsfig,multicol,bbm,simplewick}

\usepackage[usenames]{color}
\usepackage{fix-cm}

\usepackage{booktabs}

\newcommand{\Feyn}[1]{#1\kern-0.45em/}

\usepackage[english]{babel}
\usepackage{amsmath,amssymb,amsbsy,amstext, amsthm}
\usepackage{graphicx}
\usepackage{amsfonts}
\usepackage{amssymb}
\usepackage{subfigure}
\usepackage{color}
\usepackage{float}
\usepackage[small]{caption}

\makeatletter
\newlength{\apb@width}
\newcommand{\autoparbox}[2][c]{\settowidth{\apb@width}{#2}\parbox[#1]{\apb@width}{#2}}
\newcommand{\includegraphicsbox}[2][]{\autoparbox{\includegraphics[#1]{#2}}}
\makeatother

\numberwithin{equation}{section}

\def\beq{\begin{equation}}
\def\eeq{\end{equation}}

\def\bea{\begin{eqnarray}}
\def\eea{\end{eqnarray}}

\def\beq{\begin{equation}}
\def\eeq{\end{equation}}
\def\bea{\begin{eqnarray}}
\def\eea{\end{eqnarray}}

\DeclareRobustCommand{\SkipTocEntry}[4]{}

\begin{document}

\begin{titlepage}

\setcounter{page}{1} \baselineskip=15.5pt \thispagestyle{empty}

%\begin{flushright}
%{\footnotesize ITEP-TH-17/08}\\
%{\footnotesize SU-ITP-08/19}\\
%{\footnotesize SLAC-PUB-13365}\\
%{\footnotesize PUPT-2277}
%\end{flushright}
%\vfil

\bigskip\
\begin{center}
%{\fontsize{17}{30}\selectfont  \bf Desensitizing Inflation from the UV}
{\fontsize{17}{30}\selectfont  \bf Desensitizing Inflation from the Planck Scale}
%{\fontsize{17}{30}\selectfont  \bf Sequestering Inflation from the UV}
%{\fontsize{17}{30}\selectfont  \bf Reducing the UV-Sensitivity of Inflation\\ by Sequestering} %RG Flow}
\end{center}

\vspace{0.5cm}
\begin{center}
{\fontsize{13}{30}\selectfont  Daniel Baumann and Daniel Green}
\end{center}

%\vspace{0.2cm}

\begin{center}
\vskip 8pt
\textsl{School of Natural Sciences,
 Institute for Advanced Study,
Princeton, NJ 08540}

\end{center} %\vfil

%\vspace{0.8cm}

\vspace{1.2cm}
\hrule \vspace{0.3cm}
{ \noindent \textbf{Abstract} \\[0.2cm]
\noindent
A new mechanism to control Planck-scale corrections to the inflationary eta parameter is proposed. 
A common approach to the eta problem is to impose a shift symmetry on the inflaton field.
However, this symmetry has to remain unbroken by Planck-scale effects, which is a rather strong requirement on possible ultraviolet completions of the theory.
In this paper, we show that the breaking of the shift symmetry by Planck-scale corrections can be systematically suppressed if
the inflaton field interacts with a conformal sector.
The inflaton then receives an anomalous dimension in the conformal field theory, which leads to sequestering of all dangerous high-energy corrections. We analyze a number of models where the mechanism can be seen in action. 
In our most detailed example we compute the exact anomalous dimensions via $a$-maximization and show that the eta problem can be solved using only weakly-coupled physics.}
 \vspace{0.3cm}
 \hrule

\vspace{0.6cm}
%\vfil
%\begin{flushleft}
%\small \today
%\end{flushleft}

\end{titlepage}

\tableofcontents

\newpage

\section{Introduction}

A central challenge of inflation is its sensitivity to ultraviolet (UV) physics. 
Successful inflation requires that the inflaton mass is much smaller than the Hubble scale, {\it i.e.}~that the parameter 
\beq
\eta = M_{\rm pl}^2 \frac{V''}{V} \simeq \frac{m_\phi^2}{3H^2}
\eeq
 is much less than unity. The eta problem refers to the fact that UV corrections tend to drive $\eta$ to large values.
Generically, even Planck-suppressed corrections to the inflaton potential can be important and have to be understood. 
This is the main motivation for pursuing a realization of inflation in a UV-complete framework like string theory. In that case, the relevant contributions can either be computed directly, or their absence can be given a natural explanation (see \cite{McAllister:2007bg, Burgess:2007pz, Baumann:2008aq, Baumann:2009ni}
for recent reviews of inflation in string theory).
In this paper we instead propose a mechanism that solves the eta problem
 purely in effective field theory.
By coupling the inflaton field to a conformal sector with suitable properties, we show that renormalization group (RG) flow naturally suppresses all dangerous UV corrections to the inflaton self-interactions.\footnote{The effects of RG flow on inflation and its observable signatures has been previously studied (see {\it e.g.}~\cite{Stewart:1996ey,Stewart:1997wg, Lyth:1998xn}).  However, to our knowledge, the role of RG effects in solving the eta problem was not studied systematically.} 
We consider this a dynamical, low-energy solution to the eta problem.

\vskip 4pt
If the inflaton field respects an approximate shift symmetry in the infrared (IR)\footnote{By the IR we here mean a scale close to the inflationary energy scale (see Fig.~1).} then the eta problem is significantly ameliorated~\cite{Freese:1990rb}. However, even in that case, one has to worry that higher-dimension operators in the UV break the shift symmetry and lead to large corrections.  Without knowledge of the UV-completion, one should not assume that irrelevant operators respect the shift symmetry.  Moreover, if the shift symmetry has an associated conserved charge, then black hole evaporation suggests that the UV-completion should indeed break the symmetry \cite{Kallosh:1995hi,Kamionkowski:1992mf, Holman:1992us, Kamionkowski:1992ax}.\footnote{A famous consequence of the black hole no-hair theorem is the fact that the global charge of a black hole is not defined. Consider therefore a scattering process which involves an initial state with definite global charge and an intermediate state of virtual black holes. The black holes evaporate into a a final state with different global charge indicating that the global symmetry is broken. At sub-Planckian energies the symmetry breaking is described by higher-dimensiuon operators in the effective theory. Standard Wilsonian thinking then suggests that the higher-dimension operators are suppressed by appropriate powers of the Planck mass, and that all dimensionless couplings are of order one.}  It has been suggested that these arguments can be avoided if
the global symmetry is gauged in the UV.  This possibility has been realized in string theory \cite{Eva, Liam, Berg:2009tg} and in extra-dimensional constructions \cite{ArkaniHamed:2003mz, Kaplan:2003aj}, but its validity ultimately still requires knowledge of the UV-completion.
Here we will take a different approach: we will show how to control shift symmetry breaking UV operators. 
Our models will have an approximate symmetry in the IR that is badly broken by non-renormalizable corrections in the UV, but RG flow naturally suppresses the couplings of all dangerous operators. In this way, a rather generic effective field theory in the UV flows to a theory with small eta parameter in the IR.

Similar ideas have been put forward in particle physics to explain hierarchies that are much more severe than the hierarchy between the inflaton mass and the Hubble scale during inflation (typically, particle physicists use RG flow to explain hierarchies as large as $10^{-32}$~\cite{Susskind:1978ms, Weinberg:1975gm, Witten:1981nf}
and no smaller than $10^{-3}$~\cite{Nelson:2000sn}).  
For instance, in models of supersymmetry breaking, Planck-suppressed operators can give rise to flavor-violating soft masses $m^2_{\rm FV}$. Unless generic UV contributions are suppressed, the typical size of $m_{\rm FV}^2$ is inconsistent with experiment.  In `conformal sequestering'~\cite{Luty:2001jh,Luty:2001zv} flavor-blind supersymmetry breaking is achieved via a mechanism that is completely analogous to the mechanism that we propose for inflation in this paper. By coupling the supersymmetry breaking fields to a conformal sector, operators transmitting supersymmetry breaking can receive large anomalous dimensions and the flavor-violating masses ($m^2_{\rm FV}$) are suppressed relative to the flavor-diagonal soft masses ($m^2_{\rm FD}$). However, in that case, the suppression has to be by a very large factor, $\frac{m^2_{\rm FV}}{m^2_{\rm FD}} < 10^{-6}$, while for inflation the required effect is rather mild, $\eta = \frac{m^2_\phi}{3H^2} \lesssim 10^{-2}$.
This lets us hope that our mechanism can operate at weak coupling using only perturbative physics, while particle phenomenology is driven to the strong coupling regime.

Our basic idea is very simple and we believe it to apply to a wide range of inflationary mechanisms.
For concreteness, however, we will study a specific supersymmetric model in detail. In that case there will be dangerous dimension-five and dimension-six operators in the K\"ahler potential which we control be a combination of a discrete $\mathbb{Z}_2$ symmetry and conformal sequestering.
Our main motivation for writing down the specific model in \S\ref{sec:susy} is to give an `existence proof' of the sequestering mechanism, while allowing for exact computations. However, since the basic idea of our paper is very intuitive, we believe that simpler (but maybe less exact or controlled) examples should exist.

\vskip 4pt

The plan of the paper is as follow:
In \S\ref{sec:outline} we review the eta problem and explain the basic mechanism by which we aim to suppress dangerous operators. 
Next, in \S\ref{sec:susy}, we introduce a concrete supergravity model in which i) the problem is apparent and ii) the solution can be discussed in detail.
We compute the anomalous dimension of the inflaton field induced by coupling the theory to $SU(N_c)$ Yang-Mills with $3 N_c >  N_f > 3 N_c /2$ flavors. We show that our idea is safely in the perturbative regime.
In \S\ref{sec:linear} we illustrate that the idea isn't guaranteed to work by explaining its failure for a theory with linear superpotential. 
We summarize the necessary conditions that are required for a successful decoupling of dangerous operators.
In \S\ref{sec:nonSUSY} we extend our considerations to non-supersymmetric models, where
we discuss the renormalization of curvature couplings in detail.
In \S\ref{sec:gauged} we contrast our models with models involving gauged symmetries in the UV.
We make some concluding remarks in \S\ref{sec:conclusion}.

\newpage
\section{Outline of the Basic Idea}
\label{sec:outline}

\subsection{A New Solution to the Eta Problem}

In the absence of any specific symmetries protecting the inflaton potential, integrating out Planck-scale degrees of freedom generically adds the following contribution to the Lagrangian
\beq
\label{equ:O6}
 \frac{{\cal O}_6}{M_{\rm pl}^2} = \frac{{\cal O}_4}{M_{\rm pl}^2} \phi^2\, .
\eeq
If the dimension-four operator ${\cal O}_4$ has a vacuum expectation value (vev) comparable to the inflationary energy density, 
\beq
\label{equ:O4}
\langle {\cal O}_4 \rangle = c\, V\, , \quad {\rm where} \quad c \sim {\cal O}(1)\, ,
\eeq 
then this term corrects the inflaton mass by order $H \sim \frac{\sqrt{V}}{M_{\rm pl}}$, or equivalently corrects the eta parameter $\eta = M_{\rm pl}^2 \frac{V''}{V} \simeq \frac{m_\phi^2}{3H^2}$ by order one, $\Delta \eta \simeq c \sim 1$.
In supersymmetric theories the dimension-six operator in (\ref{equ:O6}) arises from the $e^{K/M_{\rm pl}^2}$ prefactor of the scalar potental if the K\"ahler potential is canonical, $K=\phi \phi^\dagger$ (see \S\ref{sec:EtaSUGRA}). The term in (\ref{equ:O6}) also arises from a non-minimal coupling to gravity, $\xi \phi^2 R$, after transforming to Einstein frame (see \S\ref{sec:nonMin}).
In addition, there may be dimension-five operators of the form
\beq
\label{equ:O5}
\frac{{\cal O}_5}{M_{\rm pl}}= \frac{{\cal O}_4}{M_{\rm pl}} \phi
\eeq
that would lead to large contributions to the first slow-roll parameter $\varepsilon = \frac{1}{2} M_{\rm pl}^2 \bigl(\frac{V'}{V} \bigr)^2$. However, since the operator in (\ref{equ:O5}) can be forbidden by a discrete $\mathbb{Z}_2$ symmetry it seems less of an immediate concern.

The most popular way to address the dangerous dimension-six operators is to assume that the inflaton respects a sufficiently powerful symmetry.
For instance, let $\phi$ be a complex field and identify the inflaton $\varphi$ with the phase of $\phi$.
Assume further that the action respects a global $U(1)$ symmetry $\phi \to e^{i \alpha}\phi$---meaning that the inflaton respects a shift symmetry $\varphi \to \varphi + \alpha$---which is only weakly broken by the inflaton potential $V(\varphi)$. It then seems that the dangerous coupling in (\ref{equ:O6}) can be forbidden.
However, rather general arguments \cite{Kallosh:1995hi} suggest that a generic theory of quantum gravity doesn't allow continuous global symmetries, so forbidding the operators (\ref{equ:O6}) in the UV is a strong requirement of a possible UV-completion (a requirement that recently has been met in string theory, {\it e.g.}~\cite{Eva, Liam, Berg:2009tg}).

In this paper we would like to pursue a different idea: we will allow the most general set of symmetry breaking operators in the UV, but couple the inflaton to a conformal sector which sequesters the dangerous terms.
Specifically, couplings to CFT operators ${\cal O}_c$ of the schematic form
\beq
f(\phi, \phi^\dagger)\, {\cal O}_{c}\,
\eeq
lead to wavefunction renormalization of the inflaton field
\beq
\label{wave}
Z\Bigl(\frac{\mu}{\Lambda}\Bigr) \partial_\mu \phi \partial^\mu \phi^\dagger\, ,
\eeq
where $\mu$ is the renormalization scale and $\Lambda$ is the UV cutoff of the CFT. 
This induces an anomalous dimension for the inflaton 
\beq
\label{equ:gamma}
\gamma \equiv \frac{1}{2} \frac{d \log Z}{d \log \mu}\, .
\eeq
In conformal theories, $\gamma$ is constant and (\ref{equ:gamma}) can be integrated
\beq
Z = \left( \frac{\Lambda}{\mu} \right)^{2 \gamma}\, .
\eeq
Inflation is studied most easily in the `physical' basis in which the fields are kept canonically-normalized $\sqrt{Z} \phi \to \phi$. In this basis, couplings in the potential run according to the anomalous dimensions of the associated operators, while kinetic terms are RG invariant. 

\begin{figure}[h!]
    \centering
        \includegraphics[width=0.37\textwidth]{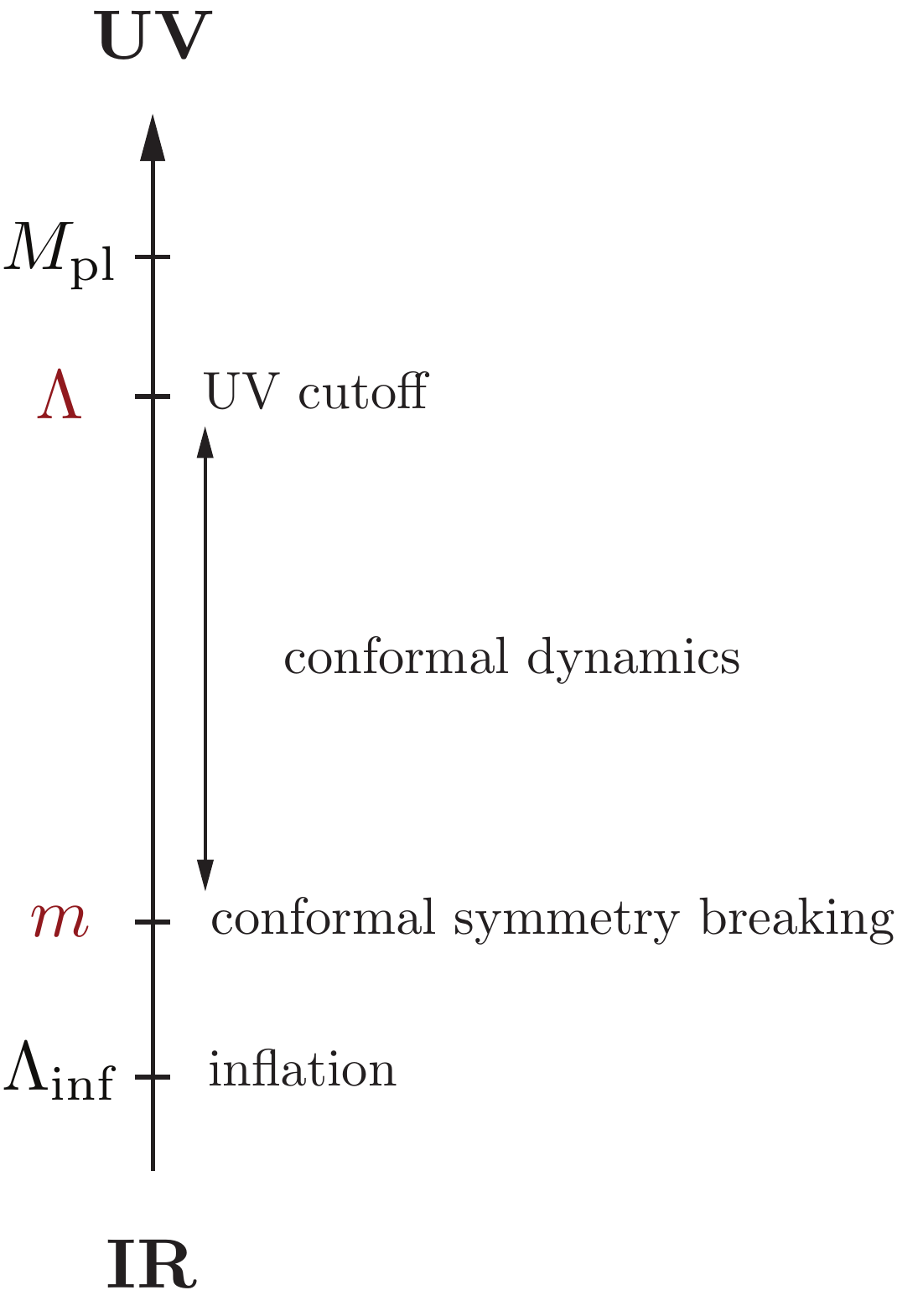}
    \caption{\sl Scales and Dynamics. In general, $M_{\rm pl}$-suppressed corrections to the action do not decouple from the inflationary dynamics at $\Lambda_{\rm inf}$. In this paper we show that this problem can be solved if the inflaton field couples to a conformal sector. In that case, RG flow in the conformal window, $m < \mu < \Lambda$, can suppress the dangerous corrections.}
    \label{fig:scales}
\end{figure}

In most cases we will assume that the conformal symmetry is broken at a scale $m$ above the inflationary scale $\Lambda_{\rm inf}$ (see Fig.~\ref{fig:scales}), so the only effect of the running is a renormalization of the IR coupling constants.
For example, the coupling in (\ref{equ:O4}) runs as follows
\beq
c(\Lambda_{\rm inf}) = \left( \frac{m}{\Lambda} \right)^{2\gamma} c(\Lambda) \, .
\eeq
If the anomalous dimension $\gamma$ is positive, the IR coupling at the inflationary scale $c(\Lambda_{\rm inf})$ is suppressed and the contribution of (\ref{equ:O6}) to $\eta$ is small even if the UV coupling $c(\Lambda)$ is order unity,
\beq
\Delta \eta = c(\Lambda_{\rm inf}) \sim \left( \frac{m}{\Lambda} \right)^{2\gamma}  \ll 1\, .
\eeq
More than 60 $e$-folds of inflationary expansion and a nearly scale-invariant spectrum of fluctuations requires $\eta \lesssim 10^{-2}$ or
\beq
\gamma \ \gtrsim \ \frac{0.1}{1 - \frac{1}{10} \log\left[ (\frac{m}{\Lambda})/10^{-10}\right]}\, .
\eeq
This suggests that an anomalous dimension as small as $\gamma \sim 0.1$ may be sufficient 
to suppress corrections to eta and we
can hope for a solution in the perturbative, weakly-coupled regime. In \S\ref{sec:susy}, we will provide an explicit example for which we are able to compute the anomalous dimension $\gamma$ directly and confirm this expectation.

\subsection{General Remarks}

Before discussing details, we would like to 
outline basic guidelines both for what qualifies as a low-energy ``solution" to the eta problem and what is necessary to achieve it.

A realistic concern in field theory solutions to the eta problem is that the necessary ingredients cannot arise in any UV-complete theory.  There are two examples that seem to fall into this category:  models with Fayet-Iliopoulos (FI) terms ({\it e.g.}~\cite{Binetruy:1996xj}) and models with large axion decay constants $f_{a} > M_{\rm pl}$ ({\it e.g.}~\cite{Freese:1990rb}).  In the absence of gravity, there seems to be no obstruction to writing down either field theory and the models seem perfectly well-defined.  
Once we include the effects of quantum gravity, however, large axion decay constant seem problematic~\cite{Kallosh:1995hi} and in fact don't seem to arise in string theory \cite{Svrcek:2006yi}.
 The status of FI terms was similar until recently, when~\cite{Komargodski:2010rb} found inconsistencies in coupling the FI term to supergravity.

Although one can always question whether a particular field theory coupled to gravity has a UV-completion, our goal will be to construct models where the field theory requires nothing more than structures known to appear in the Standard Model.  Specifically, we will demand that the small parameters of the model are radiatively stable and technically natural.  We will spend more time on supersymmetric models only because in that case these requirements seem easier to achieve.  We will allow the renormalizable Lagrangian to have continuous global symmetries, either approximate or exact.  However, these symmetries will only be accidental as we will assume that they are not respected by non-renormalizable corrections arising from the UV-completion.  We will allow for discrete $\mathbb{Z}_2$ symmetries that are respected by the irrelevant operators of the theory.  All other Planck-suppressed irrelevant operators and curvature couplings should be included with order one coefficients.  By our standards, the existence of a UV-completion for such a model is plausible enough that we will consider this a low-energy solution to the eta problem (see \S\ref{sec:comments} for further discussion of this point).
To find actual UV-completions that realize our idea would of course still be very interesting.

Applying these standards to known models, we will find that the following three 
ingredients are necessary to solve the eta problem by field theory dynamics:  
\begin{itemize}
\item[-] First, we will require that the inflaton arises as the Pseudo Nambu-Goldstone Boson (PNGB) of an approximate global symmetry (we will only need this to be a $U(1)$ symmetry and will not discuss the obvious generalizations).  This will be needed to forbid corrections to the potential that are tied to the kinetic terms and that therefore cannot be altered by RG flow.  
\item[-] Secondly, we will require a (gauged) $\mathbb{Z}_2$ symmetry that forbids possible $U(1)$ breaking dimension-five operators. We should note that this discrete symmetry is not strictly necessary.  If our anomalous dimensions are larger than unity, these terms would also be suppressed by RG flow.  We will not consider this possibility as we would like to work with perturbative theories, rather than moving the eta problem from uncertainties in the UV-completion to uncertainties about a strongly-coupled field theory.
\item[-] Finally, we will couple the inflaton to a CFT in such a way that its anomalous dimension is large enough to suppress the dimension-six operators by at least a factor of $10^{-2}$.  These couplings may not be arbitrary as they should not reintroduce the $U(1)$ breaking operators. 
In other words, the couplings have to be chosen in such a way that all dangerous couplings flow to zero rather than some finite value.
\end{itemize}

Sections 3--5 will contain various examples that help illustrate the role of each of these ingredients.  These sections are modular and need not be read sequentially to understand the general framework.  In Section~3, we will describe a concrete supersymmetric example that realizes all of the above requirements.  This example illustrates that the above conditions are sufficient to build a working model.  In Section 4, we will discuss why supersymmetric models with a linear superpotential cannot be saved using RG flow.  While this does not prove that all our requirements are absolutely necessary, it should illustrate the host of problems that can arise in models where one tries to relax these conditions.  In Section 5, we will discuss generalizations to non-supersymmetric field theories.  The same requirements and solutions exist in non-supersymmetric models with only the added difficulty of constructing a radiatively stable model in the first place.

\section{A Representative SUSY Model}
\label{sec:susy}

We now flesh out these ideas by constructing an explicit field theory model in which the sequestering effect occurs and is computable in detail.

\vskip 6pt
In \S\ref{sec:EtaSUGRA} and \S\ref{sec:Nima}, we review the eta problem in supergravity and show how it appears to be solved in a model with a shift-symmetric K\"ahler potential. 
We then explain that generic K\"ahler corrections spoil this success: $M_{\rm pl}$-suppressed dimension-five and dimension-six operators induce large corrections to $\varepsilon$ and $\eta$, respectively.
We therefore, in \S\ref{sec:mod}, introduce a slight modification of the model where a $\mathbb{Z}_2$ symmetry deals with the dangerous dimension-five terms. 
We couple this theory to a CFT in such a way that no new dangerous couplings are induced (\S\ref{sec:nonR}) and that conformal sequestering suppresses the dimension-six K\"ahler corrections (\S\ref{sec:seq}).
In \S\ref{sec:aMax}, we compute the {\it exact} anomalous dimensions of all fields and prove that small eta can be achieved in a controlled way at weak coupling.
We make further comments about the benefits of the weak coupling regime in \S\ref{sec:weak}.
Finally, in \S\ref{sec:comments} we summarize the main features of the model and discuss the UV-completion of approximate symmetries in the superpotential.

\subsection{The Eta Problem in Supergravity}
\label{sec:EtaSUGRA}

We begin by reviewing the supergravity version of the eta problem \cite{Copeland:1994vg}.
In supergravity the potential for a scalar field $\phi$ is
\beq
\label{equ:F}
V = e^{K/M_{\rm pl}^2} \left[ K^{\phi \bar \phi} D_\phi W \overline{D_\phi W} - \frac{3}{M_{\rm pl}^2} |W|^2 \right]\ ,
\eeq
where $K(\phi, \phi^\dagger)$ and $W(\phi)$ are the K\"ahler potential and the superpotential, respectively, and
\beq
D_\phi W \equiv \frac{\partial W}{\partial \phi} + \frac{W}{M_{\rm pl}^2} \frac{\partial K}{\partial \phi}\, .
\eeq
During inflation a source of vacuum energy breaks supersymmetry and induces an F-term of some field $X$: $F^2_X = \sigma^4$.
The inflaton $\phi$ then automatically receives a mass from couplings in (\ref{equ:F}),
\beq
\label{VF}
V = e^{K(\phi, \phi^\dagger)/M_{\rm pl}^2} \Bigl[ \sigma^4 + \cdots \Bigr] = \sigma^4 \Bigl[ 1 + K_{\phi \bar \phi} \frac{\phi \phi^\dag}{M_{\rm pl}^2} + \cdots  \Bigr] \, .
\eeq
For a canonical K\"ahler potential, $K(\phi, \phi^\dagger) = \phi^\dagger \phi$, this results in an order unity contribution to the eta parameter 
\beq
\label{eta}
\eta = 1 + \cdots
\eeq
The $\dots$ in (\ref{VF}) and (\ref{eta}) stand for terms that depend on the precise structure of the K\"ahler potential and the superpotential. 
It is conceivable that these omitted terms can ameliorate the eta problem. For instance, if the superpotential is linear in the inflaton $\phi$ it leads to contributions that cancel the dangerous dimension-six operator. Even for more general forms of the superpotential it is possible that an accidental cancellation between competing terms allows small $\eta$ (at least for a finite range of $\phi$). Nevertheless, to prove that fine-tuning is indeed an option typically still requires knowledge of the UV-completion. This type of one part in a hundred fine-tuning of the eta parameter has been pursued recently in many string theory models, {\it e.g.}~\cite{Baumann:2010sx, Krause:2007jk, Burgess:2006cb}.

Alternatively, a promising approach to realize a technically natural small value for $\eta$ is to make the inflaton a Goldstone boson with small mass protected by a shift symmetry.
Consider, for example, a superpotential which spontaneously breaks a global $U(1)$ symmetry
\beq
\label{equ:U1}
W = S(\Phi \bar \Phi - f^2)\, ,
\eeq
where $\Phi$ and $\bar \Phi$ are independent chiral superfields whose bottom components are the scalar fields $\phi$ and $\bar \phi$. Here and in the following, the overbars do {\it not} denote complex conjugation.
Let the expectation values of the fields be
\beq
\Phi = f e^{\theta/f}\quad {\rm and} \quad \bar \Phi =fe^{-\theta/f}\, ,
\eeq
where $\theta=\rho + i \varphi$ is a complex scalar field.\footnote{In the following we use $\theta$ both for the chiral superfield and its bottom component. Which is meant should be clear from the context.}
The canonical K\"ahler potential then becomes
\beq
\label{equ:K}
K = \Phi^\dagger \Phi + \bar \Phi^\dagger \bar \Phi 
= 2 f^2 \cosh \left( \frac{\theta + \theta^\dagger}{f}\right)\, ,
\eeq
and the supergravity potential for $\theta$ is
\beq
V = \exp \left[2 \frac{f^2}{M_{\rm pl}^2} \cosh \frac{\theta + \theta^\dagger}{f} \right] \left[ \sigma^4 + \dots \right] \, .
\eeq
We notice that only the real part of $\theta$ acquires a mass; the shift symmetry of the Goldstone boson 
is protecting the imaginary component.
This looks like a nice solution to the eta problem; however, it {\it assumes} that shift symmetry breaking contributions in the UV are small---{\it i.e.}~we have to assume that there are {\it no} non-trivial corrections to (\ref{equ:K}).  However, generic UV-completions 
are expected to break continuous global symmetries \cite{Kallosh:1995hi}, so symmetries of the K\"ahler potential are not believed to persist beyond leading order.  In this paper we therefore relax this assumption.
We are then obliged to worry about Planck-suppressed corrections 
of the form
\beq
\label{dim6}
\Delta K\ \supset\ c \, \frac{\Phi^2}{M_{\rm pl}^2} X^\dag X\, .
\eeq
Assuming order one coupling in the UV, $c(M_{\rm pl})\sim 1$, this leads to a large contribution to the inflaton mass if the coupling stays large in the IR, $c(\Lambda_{\rm inf})\sim 1$.
%In this paper, w
We now explore whether RG flow of the coupling $c(\mu)$ can solve this problem.

\subsection{Review of Pseudonatural Inflation}
\label{sec:Nima}

For purposes of illustration, we will consider the specific supergravity model of Arkani-Hamed {\it et al.}~\cite{ArkaniHamed:2003mz} (see also \cite{Kaplan:2003aj}).
Ultimately, we will construct a slight deformation of their model which controls UV corrections to the K\"ahler potential by a combination of a discrete $\mathbb{Z}_2$ symmetry and conformal sequestering.\footnote{In the extra-dimensional construction of~\cite{ArkaniHamed:2003mz}, the global symmetry becomes a gauge symmetry thus making it possible to control some dangerous K\"ahler corrections. We comment further on this possibility in \S\ref{sec:gauged}.} First, however, we describe the model in its original form.

\subsubsection{The Original Model}

The superpotential is~\cite{ArkaniHamed:2003mz, Kaplan:2003aj}
\beq
W = \lambda_0 S(\phi_1^2+\phi_2^2 - f^2) + \frac{\lambda_1}{2} \phi_1 \psi^2 + \lambda_2 X (\psi^2 -v^2)\ ,
\eeq
where $\lambda_1^2 f^2 > 2 \lambda_2^2 v^2$.
The first term in $W$ 
is the same as in (\ref{equ:U1}) if we make the identifications
\begin{align}
\Phi &\equiv \phi_1 + i \phi_2 = (f+ \rho) e^{i \varphi/f} \, ,\\
\bar \Phi &\equiv \phi_1 - i \phi_2 = (f - \rho) e^{- i \varphi/f} \, ,
\end{align}
where $\phi_1$ and $\phi_2$ are complex fields.
 This term preserves a $U(1)$ symmetry which is spontaneously broken. As before, the Goldstone boson $\varphi$ associated with the broken symmetry will be the inflaton. 
 Without loss of generality, we assume that the flat modulus $\rho$ is stabilized at $\rho \equiv 0$ after supersymmetry breaking. 
 The second term in $W$ breaks the $U(1)$ explicitly and gives the Goldstone mode a potential. 
 The field $\psi$ is the standard waterfall field of hybrid models of inflation~\cite{LindeHybrid}. During inflation it is stabilized at $\psi=0$. Finally, the last term in $W$ includes the field $X$ whose F-term dominates the inflationary potential energy, $V_0 \approx |F_X|^2 = \lambda_2^2 v^4$.
The K\"ahler potential takes the same form as in~(\ref{equ:K}). In particular, it respects the $U(1)$ symmetry.
Given this input (and for now assuming no other contributions to $W$ and $K$), the inflationary potential receives two main contributions:
\begin{enumerate}
\item[i)] a {\it loop-suppressed supergravity coupling} \cite{Kaplan:2003aj}
\beq
\label{equ:V1}
\delta K = \frac{\bar \lambda_1^2}{16\pi^2} (\Phi^\dag \bar \Phi + h.c.) \qquad \Rightarrow \qquad V_1 = V_0 \Bigl( 1 - \frac{\bar \lambda_1^2}{4 \pi^2} \frac{f^2}{M_{\rm pl}^2} \sin^2 \frac{ \varphi}{f}  \Bigr) \, ,
\eeq
where $\bar \lambda_1^2 \equiv \lambda_1^2 \log(\frac{\Lambda}{f})$ and we dropped a small constant term, $V_0(1 + \frac{\bar \lambda_1^2}{8\pi^2} \frac{f^2}{M_{\rm pl}^2}) \approx V_0$. 
\item[ii)] a {\it one-loop Coleman-Weinberg contribution} 
\beq
\label{equ:V2}
V_2 = V_0 \frac{\lambda_2^2}{4\pi^2} \log \Bigl( \frac{ \lambda_1 \cos (\varphi/f)}{\mu/f} \Bigr)\, ,
\eeq
where $\mu$ is the renormalization scale.
\end{enumerate}
The complete inflaton potential hence is
\beq
\label{pot}
V = V_0 \left( 1 -  \frac{\bar \lambda_1^2}{4 \pi^2} \frac{f^2}{M_{\rm pl}^2}\sin^2 (\varphi/f) +  \frac{\lambda_2^2}{4\pi^2} \log \bigl(\cos (\varphi/f)  \bigr)\right)\, ,
\eeq
where have absorbed small constants into $V_0$. 
%where $V_0 \equiv \lambda^2_2 v^4$.
Small $\varepsilon$ and $\eta$ can be achieved with $\bar \lambda_1 \lesssim 1$, $\lambda_2 \ll1$ and $f \ll M_{\rm pl}$~\cite{ArkaniHamed:2003mz}.  
This is easily seen  from (\ref{pot}) for the regime $\varphi \ll f$: in this case we find
\begin{align}
\eta &\simeq - \frac{\bar \lambda_1^2}{2\pi^2} - \frac{\lambda_2^2}{4\pi^2} \frac{M_{\rm pl}^2}{f^2} \, , \label{equ:eta}
\\
\varepsilon &\simeq \eta^2 \frac{\varphi^2}{M_{\rm pl}^2}\, , \label{equ:eps}
\end{align}
and inflation with $\eta \lesssim 10^{-2}$ therefore requires 
\beq
\bar \lambda_1 \lesssim 1 \qquad {\rm and} \qquad \lambda_2 \lesssim \frac{f}{M_{\rm pl}} \ll 1\, .
\eeq
 Supersymmetry makes the small value of $\lambda_2$ technically natural~\cite{ArkaniHamed:2003mz}.
From Eqn.~(\ref{equ:eps}) we infer that the model has very small epsilon parameter.
Together with the normalization 
 of the power spectrum of curvature fluctuations $\zeta$~\cite{BaumannTASI}
\beq
\Delta_\zeta^2 \equiv \frac{k^3}{2\pi^2} P_\zeta  \simeq \frac{1}{24 \pi^2} \frac{V_0}{M_{\rm pl}^4} \frac{1}{\varepsilon} \sim 10^{-10}\, ,
\eeq
this implies that the energy scale of inflation $V_0$ is very low.
We will exploit this when we consider the RG flow of couplings from the Planck scale to the inflationary scale (see Fig.~\ref{fig:scales}).
%For completeness, we note that the spectrum of fluctuations can be either red ($n_s < 1$) or blue ($n_s > 1$), depending on the sign of $\eta$ (which depends on whether $\lambda_1$ or $\lambda_2$ dominates in Eqn.~(\ref{equ:eta})),
%\beq
%n_s -1 = 2 \eta - 6\varepsilon \approx 2 \eta\, .
%\eeq

\subsubsection{Generic K\"ahler Corrections}
\label{sec:kahler}

As before, we are worried about dangerous K\"ahler potential corrections which threaten to spoil the slow-roll success of the potential (\ref{pot}).
In the present example these are {\it dimension-five operators} of the form
\beq
\label{dim5}
c_i \,  \phi_i \frac{X^\dagger X}{M_{\rm pl}} + c.c.
\eeq
and {\it dimension-six operators} of the form
\beq
\label{dim6b}
c_{ij}\, (\phi_i \phi_j + \phi_i \phi_j^\dagger) \frac{X^\dagger X}{M_{\rm pl}^2} + c.c.
\eeq
The operators in (\ref{dim5}) can be forbidden by a discrete $\mathbb{Z}_2$ symmetry (see \S\ref{sec:mod}), but the operators in (\ref{dim6b}) have to be treated seriously (see \S\ref{sec:nonR}--\S\ref{sec:seq}).
Below we will construct an explicit model where the couplings of all dangerous operators---$c_{ij}$ in (\ref{dim6b})---indeed flow to zero in the presence of suitable interactions between the inflaton sector and a hidden conformal sector.

\subsection{A Minimal Extension of the Model}
\label{sec:mod}

With the above arguments in mind, we will consider a slight modification of the model of Arkani-Hamed~{\it et al.}:
We make the model $\mathbb{Z}_2$-symmetric to forbid the dimension-five K\"ahler correction (\ref{dim5}) and couple it to a conformal sector to sequester the dimension-six K\"ahler correction (\ref{dim6b}).

\subsubsection{Discrete $\mathbb{Z}_2$ Symmetry}
\label{sec:Z2}

The upgraded superpotential is
\beq
\label{equ:WW}
W =  \lambda_0 S(\phi_1^2+\phi_2^2 - f^2) +  \frac{\lambda_1}{2} \phi_1 \psi \bar{\psi} + \lambda_2 X (\bar{\psi}^2 -v^2) + W_{\rm CFT}\  ,
\eeq
where $W_{\rm CFT}$ will couple $\phi_1$ and $\phi_2$ to a CFT. 
When $W_{\rm CFT} = 0$, this model has a $\mathbb{Z}_2 \times \mathbb{Z}_2 $ symmetry with $\phi_{1} \to - \phi_{1}$ and $\psi \to - \psi$ under the first $\mathbb{Z}_2$ and $\phi_2 \to - \phi_2$ under the second $\mathbb{Z}_2$.  
However, to forbid the dangerous operators in (\ref{dim5})
we only need the $\mathbb{Z}_2$~symmetry that transforms all three fields simultaneously.
Using this smaller symmetry will allow more flexibility in adding additional couplings to the model.
We therefore take the K\"ahler potential to be
\beq
\label{equ:Kcorr}
K = \phi_1^{\dag} \phi_1 +  \phi_2^{\dag} \phi_2 + X^{\dag}X \left[ 1 + c_1 \frac{\phi_1 \phi_1^{\dag}-\phi_2 \phi_2^{\dag}}{M_{\rm pl}^2}+ c_2  \frac{\phi_1 \phi_2^{\dag}+\phi_2 \phi_1^{\dag}}{M_{\rm pl}^2} + c_3 \frac{\phi_1 \phi_2 + \phi_1^{\dag} \phi_2^{\dag}}{M_{\rm pl}^2} \right] \ +\ \cdots \ ,
\eeq
which is invariant under the $\mathbb{Z}_2$ with $\Phi \to - \Phi$ and $\bar \Phi \to - \bar \Phi$.
By the usual logic that the UV-completion should not preserve continuous global symmetries \cite{Kallosh:1995hi}, one should take the coefficients $c_i$ to be of order one.  The $\ldots$ in (\ref{equ:Kcorr}) include all other operators that do not contribute significantly to the inflationary potential, including the canonical K\"ahler potential terms for other fields and operators that respect the global $U(1)$ symmetry or have dimensions greater than six.

\subsubsection{Coupling to a Conformal Sector}
\label{cop}

Next, we couple this theory to a CFT in such a way that all dimension-six operators receive {\it positive} anomalous dimensions, without inducing other dangerous operators.  The simplest way to do this is to couple the inflaton to a $SU(N_c)$ gauge theory with $N_f$ flavors such that $\frac{3}{2} N_c < N_f <  3 N_c $.  In this regime, the theory is conformal~\cite{Intriligator:1995au}.  To this theory we add the superpotential
\beq
\label{Wcft}
W_{\rm CFT} = y_1 \sum_{i=1}^{N_1} \tilde{Q}_{i} Q_i \Phi + y_2   \sum_{j=N_1+1}^{N_2} \tilde{Q}_{j} Q_j \bar{\Phi} + m\sum_{i=1}^{N_1} \tilde{Q}_i {\cal Q}_{N_2+i} + m\sum_{j=N_1+1}^{N_2} \tilde{\cal Q}_{N_2+j} Q_{j}\, ,
\eeq
where $Q$ and $\tilde{Q}$ are (anti-)fundamentals of $SU(N_c)$.
Here, we have split the flavor fields into three distinct classes:
\begin{itemize}
\item {\it $N_1$ flavors that couple to $\Phi$}\, :
$$\hspace{-0.6cm} \{ Q_i, \tilde Q_i \}  \qquad i = 1, \dots , N_1 $$
To preserve the $\mathbb{Z}_2$ symmetry these fields transform as: $\{ Q_i, \tilde Q_i \} \to \{ - Q_i, + \tilde Q_i \}$.
\vskip 10pt
\item {\it $N_2-N_1$ flavors that couple to $\bar \Phi$}\, :
$$\quad \{ Q_j, \tilde Q_j \}  \qquad j = N_1+1, \dots , N_2 $$
To preserve the $\mathbb{Z}_2$ symmetry these fields transform as: $\{ Q_j, \tilde Q_j \} \to \{+ Q_j, - \tilde Q_j \}$. 
\vskip 10pt
\item {\it $N_f - N_2$ flavors that do not couple to either $\Phi$ or $\bar \Phi$}\, :
$$ \quad \{ {\cal Q}_k, \tilde {\cal Q}_k\} \qquad k = N_2+1, \dots , N_f $$
These fields transform as: $\{ {\cal Q}_k,  \tilde {\cal Q}_k\}  \to \{+ {\cal Q}_k,  +\tilde{\cal Q}_k\}$.
\end{itemize}
  The mass terms in (\ref{Wcft}) hence involve one $Q$ that couples to the $\Phi$'s and one that does not.  This split of the flavor fields may seem like an odd choice, but 
    it is simply required to ensure that the $\mathbb{Z}_2$ symmetry is unbroken. 
  In principle, we could allow the couplings to the CFT to break the $\mathbb{Z}_2$ symmetry without ruining the potential,
  but we here prefer to work with couplings that preserve the $\mathbb{Z}_2$ exactly.  
   In doing so, we have assumed that $N_f > 2 N_2$.  We will also choose $N_f = 3 N_c - k$ with the idea that our anomalous dimensions will be perturbative in a $k/N_c$ expansion (see \S\ref{sec:aMax}).  
Finally, we require that $N_2 < N_f - N_c$ in order to avoid generating a non-perturbative superpotential for $\Phi$ after integrating out the flavors.  

\subsection{Non-Renormalization}
\label{sec:nonR}

We will now argue that no dangerous $U(1)$ violating operators are generated by the CFT couplings in (\ref{Wcft}).
We will achieve this via a spurion analysis, promoting the couplings in (\ref{Wcft}) to fields that transform under the global $SU(N_f) \times SU(N_f)$ symmetry of the theory: 

Consider the CFT superpotential in (\ref{Wcft}).  We will rewrite it as follows
\beq
\label{Wcft2}
W_{\rm CFT} = y_1^{\bar{\imath} \bar{\jmath}} \tilde{Q}_{i} Q_j \Phi + y_2^{\bar{k} \bar{l} } \tilde{Q}_{k} Q_l \bar{\Phi} + m_1^{\bar{m} \bar{n} }\tilde{Q}_m Q_{n} + m_2^{\bar{p} \bar{q}} \tilde{Q}_{p} Q_{q}\, ,
\eeq
where the sum on repeated indices is implied.  To return to the form of (\ref{Wcft}), one simply specifies the matrices for each of the couplings ({\it e.g.},~$y_1^{\bar \imath \bar \jmath} = y_1 \delta^{\bar{\imath} \bar{\jmath}}$ for $i,j \leq N_1$ and 0 otherwise).  Before adding these couplings, the CFT has an $SU(N_f) \times SU(N_f)$ symmetry acting on the $Q$'s and $\tilde{Q}$'s.
 Under this symmetry, the couplings $y_1$ and $y_2$ and the mass matrices $m_1$ and $m_2$ transform as anti-fundamentals under both groups.  Any corrections induced by the couplings in (\ref{Wcft2}) should respect the global symmetry, so we should contract all $SU(N_f)$ indices with $\delta_{\bar{a}  a}$, $\epsilon_{ a_1 \ldots a_{N_f}}$ or $\epsilon_{ \bar{a}_1 \ldots \bar{a}_{N_f}}$.

We will only be interested in terms that do not involve the $Q$ fields, since terms with explicit $Q$ fields do not lead to significant contributions to the inflaton potential.  Therefore, possible corrections must be invariant under the flavor symmetry by contracting all the indices of the couplings.  In order to contract with an epsilon tensor, the flavor symmetry must be completely broken.  We could forbid such terms by coupling to less than $N_f$ flavors in the superpotential.  In any case, such corrections would appear at very high order in the coupling $y$ and thus are highly suppressed.  All terms constructed by contracting with epsilon tensors are therefore harmless.

It should be clear that corrections proportional to $\delta_{\bar{a}  a} \delta_{\bar{b}  b} \, (y_i^{\dag})^{ a b} (y_j)^{ \bar{a} \bar{b}}$ and $\delta_{\bar{a}  a} \delta_{\bar{b}  b} \, (y_i^{\dag})^{ a b} (m_j)^{ \bar{a} \bar{b}}$ are consistent with the global symmetries.  Before specifying the actual form of the matrices $y_{i}$ and $m_{i}$ in (\ref{Wcft2}), we would expect many corrections to be possible. Some of them would be dangerous. For example, $  (y^{\dag}_{1})_{a b} (y_2)^{\bar{a} \bar{b}} \Phi^{\dag} \bar{\Phi}$ is allowed by the symmetries.  Similarly, we can have corrections of the form $(y_{1}^{\dag})_{ab} (m_{1})^{\bar a \bar b} \Phi^\dag$.  Any of these corrections could alter the potential and we would have to take them into account. 

After determining the form that all corrections can take, we evaluate them with our specific choice of couplings in (\ref{Wcft}).  Notice that we have chosen the couplings such that no pair of indices is shared by different couplings.  For these choices, corrections like $(y^{\dag}_{1})_{a b} (y_2)^{\bar{a} \bar{b}} \Phi^{\dag} \bar{\Phi}$ in fact vanish.  As a result, the only non-zero corrections are proportional to $y_1^{\dag} y_1$ or $y_2^{\dag} y_2$.  
Alternatively, we could also give our couplings charge under the $U(1)$ symmetry.  Because $y_i^{\dag} y_i$ is $U(1)$ invariant, the combination of $\Phi$ fields must also be $U(1)$ invariant.  As a result, we will only generate terms of the form $\Phi^{\dag} \Phi$ (or $\bar \Phi^{\dag} \bar \Phi$), but none of the form $\Phi^2$ (or $\bar \Phi^2$).  Therefore, our CFT produces no dangerous couplings.

\vskip 4pt
The choice of couplings in (\ref{Wcft}) has the further benefit of making the anomalous dimensions of the model exactly computable via the method of `$a$-maximization'~\cite{Intriligator:2003jj} (see \S\ref{sec:aMax}).  In general, the dimensions of K\"ahler potential terms are of course not determined by $a$-maximization.  Specifically, we have so far been using a notation in which the anomalous dimensions of operators like $\phi_1^{\dag} \phi_1- \phi_2^{\dag} \phi_2$ are given by $ 2 \gamma$ where $\gamma$ is determined by the R-charges of $\phi_{1}$ and $\phi_2$.  In most CFTs, this is not the case.  Non-chiral operators have dimensions that are unrelated to the chiral dimensions.  Furthermore, there are non-chiral operators that have protected dimensions of two because they form a supermultiplet containing a conserved current.

Fortunately, we are not interested in the most general K\"ahler corrections, but only in those that affect the inflaton potential.  As we have discussed, only operators charged under the global $U(1)$ symmetry can change the potential.  For operators like $\Phi^2 X^{\dag} X$, the dimension follows from the chiral dimension because the CFT only couples to $\Phi^2$ which is chiral.  However, we still have operators like $\Phi^{\dag} \bar{\Phi} X^{\dag} X$, which involve non-chiral combinations of fields coupled to the CFT.  As with the $\Phi^2$ operators, one cannot forbid all renormalization without appealing to holomorphy.  Specifically, global symmetries allow additional contributions to the anomalous dimension of the form $ (y_1^{\dag} y_1)^n (y_2^{\dag} y_2)^m$.  Nevertheless, because distinct flavor fields couple to $\Phi$ and $\bar{\Phi}$,  these interactions factorize into a renormalization of $\Phi$ and a renormalization of $\bar{\Phi}$.  Appealing again to holomorphy, there should be no such renormalization.  One can indeed check that this is the case to all orders in perturbation theory.  This situation is similar to the case where $\Phi$ and $\bar{\Phi}$ are coupled to two different, decoupled CFTs.  As a result, the only contribution to the dimensions of these operators comes from wavefunction renormalization and may indeed be determined by $a$-maximatization (see \S\ref{sec:aMax}).

\subsection{Conformal Sequestering}
\label{sec:seq}

To get a computable example, we have added to the inflaton sector a conformal sector. 
The inflaton then develops an anomalous dimension $\gamma$ via the superpotential couplings to operators in the new sector (\ref{Wcft}).
In the `holomorphic' basis where the superpotential is not renormalized this changes the inflaton kinetic term
\beq
{\cal L} = \int d^4 \theta\, Z\Phi^\dagger \Phi\, ,
\eeq
and similiarly for $\bar \Phi$.
In the `physical' basis in which the fields are kept canonically-normalized, $\sqrt{Z} \Phi \to \Phi$, this implies that couplings in the superpotential and higher-dimension terms in the K\"ahler potential run according to the anomalous dimensions of the associated operators, while the leading terms in the K\"ahler potential are RG invariant. 

The mass terms in (\ref{Wcft}) break conformal invariance at the scale $m$.  We will usually take $m \geq f$ so that the CFT decouples above the scale of inflation (see Fig.~\ref{fig:scales}).  
In the IR, the renormalized couplings of the dimension-six K\"ahler corrections in (\ref{equ:Kcorr}) therefore are
\beq
c_i(\Lambda_{\rm inf}) = \left( \frac{m}{\Lambda} \right)^{2\gamma} c_i(\Lambda)   \, .
\eeq
If the anomalous dimension $\gamma$ is positive, then the coupling is suppressed in the IR and the contribution of (\ref{equ:Kcorr}) to $\eta$ can be small.
Above we argued that quite a small anomalous dimension---{\it e.g.},~$\gamma \sim 0.1$---is sufficient to suppress corrections to $\eta$ to acceptable levels.
Below we will confirm in detail that this mechanism can indeed solve the supergravity eta problem
 in the perturbative, weakly-coupled regime.

Below the scale of conformal symmetry breaking $m$, our model is very similar to that of \S\ref{sec:Nima}. 
In particular, the scalar potential will receive the same contributions (\ref{equ:V1}) and (\ref{equ:V2}).
However, in addition there will be contributions from the K\"ahler potential corrections
\beq
\Delta V = V_0 \frac{f^2}{M_{\rm pl}^2} \left(\frac{m}{\Lambda}\right)^{2 \gamma} \Bigl[ \tilde{c}_1 \sin^2(\varphi / f)+ \tilde{c}_2 \sin(\varphi / f) \cos(\varphi / f) \Bigr]\ ,
\eeq
where $\tilde{c}_1$ and $\tilde{c}_2$ are order one coefficients.  We have removed terms independent of $\varphi$ to reduce $c_1$, $c_2$ and $c_3$ to the two constants $\tilde{c}_1$ and $\tilde{c}_2$.
This leads to the following corrections to the slow-roll conditions for the model
\begin{align}
\Delta \varepsilon &= \frac{1}{2}
 \frac{f^4}{M_{\rm pl}^4} \left(\frac{m}{\Lambda}\right)^{4 \gamma} \Bigl[2 \tilde{c}_1 \sin(\varphi / f) \cos(\varphi /f)+ \tilde{c}_2 (1 -2 \sin^2(\varphi / f))\Bigr]^2\ , \\
\Delta \eta &= 
  \frac{f^2}{M_{\rm pl}^2} \left(\frac{m}{\Lambda} \right)^{2 \gamma} \Bigl[ 2 \tilde{c}_1(1- \sin^2(\varphi /f)) - 4 \tilde{c}_2  \cos(\varphi/f)\sin(\varphi / f) \Bigr]\ .
\end{align}

Given that the mass of the waterfall field is proportional to $\phi_1 =f \cos(\varphi/f)$, we will assume that inflation takes place when $\varphi \ll f$.  In this limit, 
\begin{align}
\Delta \varepsilon &\simeq \frac{(\tilde c_2)^2}{2}
\frac{f^4}{M_{\rm pl}^4} \left(\frac{m}{\Lambda}\right)^{4 \gamma} \, , \\ 
\Delta \eta &\simeq   2 \tilde c_1 \left(\frac{m}{\Lambda}\right)^{2 \gamma}\, . 
\end{align}
For $\tilde c_i \sim {\cal O}(1)$ we require $(\frac{m}{\Lambda})^{2 \gamma} \lesssim 10^{-2}$ to avoid large eta.
We now aim to explain this number via a concrete computation in the CFT.

\subsection{Anomalous Dimensions via $a$-Maximization}
\label{sec:aMax}

Without the coupling to 
$\Phi$ and $\bar \Phi$, the anomalous dimensions for the flavor fields are easily determined from the NSVZ beta function \cite{Novikov:1983uc}.  Because of the unbroken flavor symmetry, all the flavors have a common anomalous dimension $\widetilde \gamma_Q = - (3 N_c - N_f ) /  2 N_f$.  However, when we couple $N_2 < N_f$ flavors to 
$\Phi$ and $\bar \Phi$ as in (\ref{Wcft}), there are five different anomalous dimensions, that cannot be determined from the vanishing of the beta functions alone.  However, as we now show, they can be determined via $a$-maximization~\cite{Intriligator:2003jj}.

In general, $a$-maximization does not apply to non-chiral operators.  However, as we discussed in \S\ref{sec:nonR}, our CFT couplings are such that the only dimension-two non-chiral operators that receive corrections beyond wavefunction renormalization are those that preserve the $U(1)$ symmetry, such as $\Phi^{\dag} \Phi$ and $\bar{\Phi}^{\dag} \bar{\Phi}$. 
These operators do not directly influence the inflaton potential.  As a result, the operators of interest are only sensitive to the CFT through the dimensions of the chiral operators and so their anomalous dimensions are indeed given by $2 \gamma$, where $\gamma$ is determined by the R-charges of $\Phi$ and $\bar{\Phi}$.

Let us briefly outline the $a$-maximization procedure to compute these dimensions.  For a superconformal field theory, the dimensions of chiral operators are determined by their superconformal R-charges, $\Delta = \frac{3}{2} R \equiv 1 + \gamma$.  Thus, finding the dimensions of chiral operators is equivalent to finding these charges.   However, there may be more than one anomaly-free $U(1)_R$ symmetry, but only a single combination can form the superconformal R-symmetry that relates to dimensions.  A priori, unless there is a unique anomaly-free $U(1)_R$, one would be unable to determine the correct R-charges.  However, as was shown in \cite{Intriligator:2003jj}, the unique superconformal R-charge maximizes the superconformal anomaly coefficient $a$.  

In a four-dimensional $\mathcal{N}=1$ superconformal theory, the conformal anomaly coefficients $a$ and $c$ are determined by the superconformal R-charges.  We will be interested in $a$, which is given by\footnote{Here we have ignored an overall factor of $\frac{3}{32}$.}
\beq
a =3 \text{Tr} (R^3) - \text{Tr} (R)\, .
\eeq
This relation occurs because the R-current is related by supersymmetry to the stress tensor.  We now let the charges be given by an arbitrary linear combination of anomaly-free R-charges and determine the correct superconformal R-charge by the one that yields a local maximum for $a$.  This procedure requires the correct identification of the full set of global symmetries at the fixed point.  Because our fixed points will be perturbative,  there will be no subtlety in identifying the possible R-symmetries.  

For the problem at hand, we have five different R-charges, one for $Q_i$ ($R_i$), $Q_j$ ($R_j$), ${\cal Q}_{k}$ ($R_k$), $\Phi$ ($R_{\Phi}$) and $\bar{\Phi}$ ($R_{\bar{\Phi}}$). 
Then,
\beq
\label{a}
a = 2 (N_c^2-1) + \sum_I {\rm dim}(r_I)  \Bigl[  3(R_I-1)^3 - (R_I-1) \Bigr]\,,
\eeq
where $I$ runs over the five types of fields and $r_I$ is the representation of each field.
 These charges $R_I$ cannot vary independently because of constraints:   Imposing that the R-symmetry is anomaly-free requires that
\beq
\label{con1}
N_c  + N_1 (R_i - 1) + (N_2 - N_1) (R_j - 1) + (N_f - N_2) (R_k -1)= 0.
\eeq
Furthermore,
 the superpotential terms have R-charge 2, which implies that
\begin{align}
R_{\Phi} + 2 R_i&= 2  \, , \label{con2}\\
R_{\bar{\Phi}} + 2 R_j&= 2 \, . \label{con3}
\end{align}
With the constraints (\ref{con1}), (\ref{con2}) and (\ref{con3}), equation (\ref{a}) becomes a function of $R_\Phi$ and $R_{\bar \Phi}$\footnote{Dropping constant terms that are independent of $R_\Phi$ and  $R_{\bar \Phi}$.}
\begin{align}
a(R_\Phi, R_{\bar \Phi}) &=   3(R_\Phi-1)^3 - (R_\Phi-1) + 3(R_{\bar \Phi}-1)^3 - (R_{\bar \Phi}-1) \nonumber \\
& \hspace{0.5cm} - \frac{3}{8} N_1 R_\Phi^3  -  \frac{3}{8}(N_2-N_1) R_{\bar \Phi}^3   - 3 \frac{(N_c - \frac{1}{2}N_1 R_\Phi - \frac{1}{2}(N_2-N_1) R_{\bar \Phi} )^3}{(N_f-N_2)^2}  \, . \label{a2}
\end{align}
One can now determine the R-charges by finding the local maximum of this function.
The R-charges (and thus dimensions) can thus be determined for general $N_1$, $N_2$ and $N_f$. 
However, the general result is not very illuminating. To gain intuition from a simple analytical solution, we will consider the special case where $N_2 = 2 N_1$.
Furthermore, because we are aiming for perturbative anomalous dimensions, we will work in the {\it Banks-Zaks window}:  $N_f = 3 N_c -k $, where $k \ll N_c$.  After integrating out the massive flavors, we would like to have less than $\frac{3}{2} N_c$ flavors remaining.  Therefore, we will make the choice $N_1 =  \frac{3N_c-k}{4} = \frac{N_f}{4}$.  
In this case the exact result for the anomalous dimension $\gamma_{\Phi} = \frac{3}{2}R_\Phi - 1$ is
\beq
\label{exact}
\gamma_\Phi = \frac{8+3 N_c}{16} \left[ 1 - \sqrt{1- \frac{96 N_c}{(8+3N_c)^2} \frac{x}{3-x}}\,\right]\ , \qquad {\rm where} \qquad x \equiv \frac{k}{N_c}\, .
\eeq
Expanding this result in small $\frac{k}{N_c}$, gives
\beq
\label{approx}
\gamma_{\Phi} = \frac{N_c}{8 + 3 N_c} \left(\frac{k}{N_c}\right) + \mathcal{O}\left(\frac{k^2}{N_c^2}\right)\, .
\eeq
In the limit where $ 3 N_c \gg 8$, the leading order result is the same as in the case where one couples equally to all the flavors 
\beq
\gamma_\Phi \approx \frac{3N_c}{8+3N_c} {\widetilde \gamma}_{\Phi}\, ,
\eeq
where ${\widetilde \gamma}_{\Phi} \equiv (3 N_c - N_f ) / N_f$.
Ultimately, we expect the $k / N_c$ expansion to be related to a weak-coupling expansion (see \S\ref{sec:weak}).  We see that to get an anomalous dimension of the order of $\gamma_{\Phi} \sim \frac{1}{9}$, we will need $k \sim \frac{N_c}{3}$ for large $N_c$.  One can easily check that the difference between the answer to leading order in $k / N_c$ (\ref{approx}) and the full answer (\ref{exact}) then is $\frac{1}{72}$.  Therefore, 
there is reason to believe that the coupling is weak enough to use the loop expansion.

\subsection{Weak Coupling {\it vs.}~Strong Coupling} 
\label{sec:weak}

At one loop, we can directly calculate the anomalous dimensions for the different fields.  For the flavors that don't couple to $\Phi$ and $\bar \Phi$, the only contribution comes from the gauge coupling and is given by
\beq
\gamma_{{\cal Q}_{k}} = - \frac{g^2}{8 \pi^2} \frac{N_c^2-1}{N_c}\, ,
\eeq
while the flavors that couple to the $\Phi$'s have dimensions
\beq
\gamma_{Q_{i,j}} = - \frac{g^2}{8 \pi^2} \frac{N_c^2-1}{N_c} + \frac{y^2}{8 \pi^2}.
\eeq
Here, we are assuming the same special case as before, $N_2 = 2 N_1$, and thus $y_1 = y_2 \equiv y$ at the fixed point.  The anomalous dimensions for $\Phi$ and $\bar \Phi$ are given by
\beq
\gamma_{\Phi} = \gamma_{\bar{\Phi}} = N_c N_1 \frac{y^2}{8 \pi^2}\, .
\eeq
Again, working at large $N_c$, we find that $\gamma_\Phi \sim \frac{1}{9}$ if $\frac{g^2}{8 \pi^2} N_c \sim \frac{1}{9}$ and $\frac{ 3 N_c^2}{4} \frac{y^2}{8 \pi^2} \sim \frac{1}{9}$.

Since at large $N_c$, the anomalous dimensions of the flavors are equal up to $\frac{1}{N_c}$ corrections, the vanishing of the beta function for $g^2$ implies 
\beq
\frac{g_*^2}{8 \pi^2} \sim \frac{3 N_c - N_f}{2 N_c N_f}\, .
\eeq
Given $g_*^2$, we can determine the couplings to the inflaton at the fixed point 
\beq
\frac{y^2_*}{8 \pi^2} \sim  \frac{3 N_c - N_f}{N_1 N_c  N_f}\, .
\eeq  
Since the loop expansion is an expansion in $N_c \frac{g_*^2}{8 \pi^2} $ and $N_1 N_c \frac{y^2_*}{8 \pi^2}$, higher-loop corrections are suppressed at the fixed point.

Using these results, we can further check the relationship between weak coupling and small $\frac{k}{N_c}$.  Expanding in $k = 3 N_c - N_f \ll N_c$, we find that $N_c \frac{g_*^2}{8 \pi^2} \sim \frac{k}{6 N_c}$ and 
\beq
\gamma_\Phi = N_c N_1 \frac{y_*^2}{8 \pi^2} \sim  \frac{k}{3 N_c}\, .
\eeq  
This shows that our weak-coupling expansion at the fixed point is the same as the $\frac{k}{N_c}$ expansion---{\it cf.}~Eqn.~(\ref{approx}).  Using the $a$-maximization results, we see that the sum of higher-loop corrections to $\gamma_\Phi$ is smaller than the one-loop result by a factor of $\frac{1}{8}$.
This shows that the mechanism can operate safely at weak coupling.

\vskip 4pt
Restricting ourselves to weakly-coupled fixed points implies that we absolutely needed the additional $\mathbb{Z}_2$ symmetry in (\ref{equ:WW}) to eliminate $U(1)$ breaking dimension-five operators in the K\"ahler potential.  
If instead one was prepared to push our results into the strong coupling regime, the $\mathbb{Z}_2$ symmetry could become superfluous:
 dimension-five and dimension-six operators would both be suppressed by anomalous dimensions $\gamma_\Phi \gtrsim 1$.  
 However, if these anomalous dimensions are generated by superpotential couplings of the form $W \supset \mathcal{O}_c \Phi$, then $\Delta_{\Phi}= 1+\gamma_\Phi > 2$ requires $\Delta_{\mathcal{O}_c} < 1$.  While this seems to require that $\mathcal{O}_c$ violates the unitarity bound, the equations of motion of $\Phi$ force $\mathcal{O}_c$ to vanish in the chiral ring---{\it i.e.}~${\cal O}_c$ is not a primary operator and to which the unitarity bounds do not apply.  

This loophole has been exploited in constructing models of flavor \cite{Nelson:2000sn,Poland:2009yb}.  However, one is forced to assume that this superpotential deformation flows to an interacting fixed point rather than a massive one.  Nevertheless, if models like those in \cite{Nelson:2000sn, Poland:2009yb} are truly interacting fixed points with $\Delta_{\Phi} > 2$, we could indeed construct models that dynamically solve the eta problem without requiring any additional discrete symmetries.  We have not pursued such models here, as the uncertainty of the existence of the fixed point could be considered as severe a problem as the existence of UV-completions without the dangerous K\"ahler potential terms in the first place.

\subsection{Summary and Comments}
\label{sec:comments}

In this section, we presented a concrete supersymmetric model of inflation where the eta problem was solved dynamically. 
Inspired by \cite{ArkaniHamed:2003mz} we made the inflaton the PNGB of an approximate $U(1)$ symmetry.  The fields that were charged under the $U(1)$ were then coupled to an SQCD sector in the Banks-Zaks window.   As a result, the charged fields acquired a positive anomalous dimension.  Importantly, we assumed that the superpotential takes a special form in order to realize the approximate $U(1)$ symmetry.  Given this choice, the superpotential is radiatively stable and technically natural.  We also demanded that the model has an exact $\mathbb{Z}_2$ symmetry to forbid certain dimension-five operators in the K\"ahler potential.  
Given the specific superpotential couplings, 
the model does not have an eta problem even when the most general Planck-suppressed operators are included in the K\"ahler potential.   The couplings to the CFT forbid any dangerous $U(1)$ breaking terms from being generated and the couplings of all dangerous K\"ahler corrections flow to zero.
We calculated the anomalous dimensions directly using $a$-maximization and argued that the eta problem is solved even in the perturbative regime. 

In considering this specific model, we assumed that the superpotential with approximate $U(1)$ symmetry could be realized in some UV-completion.  
We want to make a few comments about this assumption:
because of holomorphy, any particular choice for the superpotential will of course not be modified by quantum corrections.  
The form of the superpotential is therefore technically natural and radiatively stable.
Nevertheless, it is reasonable to ask how natural the assumed structure of the superpotential is from the point of view of the UV-completion.
Specifically, it is not obvious that technical naturalness in the field theory sense implies `stringy naturalness'.
Often string compactifications come with extra constraints that are not immediately transparent from the low-energy field theory point of view. 
In the specific model being considered, we did omit certain $U(1)$ breaking terms in the superpotential. 
In field theory this is a perfectly valid thing to do since we argued that these terms are not generated by quantum corrections. Whether this approximate IR symmetry is easy to achieve in a UV-completion like string theory is largely an open question. However, our work has shifted the problem from uncomputable K\"ahler corrections to the origin of protected symmetries in the superpotential. We consider this much more amenable to concrete computations.

 In extra-dimensional UV-completions, the absence of certain terms in the superpotential can arise from locality in the extra dimensions.  However, it is well-known that moduli stabilization often interfers destructively with inflation and the preservation of IR symmetries.
 In fact, 
 in string compactifications whose volume is stabilized by a nonperturbative superpotential~\cite{KKLT} shift symmetries are often broken by superpotential interactions \cite{McAllister:2005mq, BHK, BDKMMM}.
It would be interesting to explore if this conflict with moduli stabilization persists more generally. This would imply that explicit string constructions are harder to achieve than our field theory intuition would have led us to believe.
However, we want to emphasize that without a systematic study of these issues we consider it premature to draw any such conclusion. 

Finally, let us stress that
the structure we require of our models is no different than the structure of the MSSM.
This suggests that engineering an approximate $U(1)$ in field theory can also be pursued using the Standard Model as an example. 
 By including further gauge symmetries in the model, one could hope to forbid all the global $U(1)$ breaking couplings up to dimension-four.  The approximate symmetry of the low-energy theory would then be a consequence of the gauge symmetries of the model, rather than an accident of the UV-completion.
 This can serve as a guide for how these structures could arise in explicit string constructions.
We leave a systematic exploration of these interesting questions to future work.

\newpage
\section{Failure Modes of Sequestering}
\label{sec:linear}

We shouldn't give the impression that conformal sequestering is guaranteed to solve the eta problem.  The role of the shift symmetry was important for more than just eliminating the leading supergravity contributions to eta; it also ensured that only shift symmetry breaking operators could contribute to the potential.  These terms are suppressed when the inflaton is coupled to a CFT in a way that respects the symmetry.

\vskip 4pt
In this section we show that a theory with linear superpotential (which naively is free of dangerous dimension-six operators: \S\ref{sec:linRev}) cannot be saved from UV corrections by RG flow (\S\ref{sec:linRG}).
The ways in which the model fails will teach us interesting lessons about the limits of applicability of our idea (\S\ref{sec:lessons}).
We will also explain why we believe that our approach is destined to fail for large-field models
(\S\ref{sec:large}).

\subsection{Review of the Linear Model}
\label{sec:linRev}

For a theory with canonical K\"ahler potential for a chiral superfield $\Phi$ and a superpotential $W = \sigma^2 \Phi$, the potential for $\Phi$ takes the form
\beq
V(\Phi) = e^{\Phi^{\dag} \Phi/M_{\rm pl}^2} \left( \Bigl|\sigma^2 + \frac{\Phi^{\dag} \Phi}{M_{\rm pl}^2} \sigma^2 \Bigr|^2 - 3 |\sigma|^4 \frac{\Phi^{\dag} \Phi}{M_{\rm pl}^2} \right)= \sigma^4  + \mathcal{O}\Bigl(\sigma^4 \frac{(\Phi^{\dag} \Phi)^2}{M_{\rm pl}^4} \Bigr)\, .
\eeq
Because the leading dimension-six operators cancel, this seems like a promising starting point for inflationary model-building in supergravity.  However, as with our PNGB example in \S\ref{sec:susy}, this approach does not solve the eta problem completely, as we have no good reason to forbid additional Planck-suppressed operators in the K\"ahler potential.  Specifically, if the K\"ahler potential takes the more general form 
\beq
K = \Phi^{\dag}\Phi + \frac{c_1}{M_{\rm pl}} (\Phi^{\dag} \Phi^2 + h.c.) + \frac{c_2}{M_{\rm pl}^2} (\Phi^{\dag}\Phi)^2\, ,
\eeq
then the scalar potential becomes
\beq
V(\Phi) = \sigma^4 \Bigl(1  - \frac{c_1}{M_{\rm pl}} (\Phi^{\dag} +  \Phi) + \frac{c_1^2}{2 M_{\rm pl}^2}(\Phi^{\dag} +  \Phi)^2 - \frac{c_2}{M_{\rm pl}^2} \Phi^{\dag} \Phi \Bigr) + \cdots
\eeq
For $c_{1,2} \sim \mathcal{O}(1)$, these corrections lead to order one corrections to $\eta$ and to large contributions to $\varepsilon$, just as in our previous example.  One might hope that the same technique of coupling the inflaton to a CFT could help to solve this problem here as well.  This will turn out not to be the case, but the ways in which it fails will be instructive and quite useful for understanding the possible pitfalls of these types of models.

\subsection{RG Flow of the Linear Model}
\label{sec:linRG}

Let us start by coupling the inflaton $\Phi$ to a CFT through a term in the superpotential $W \supset y \mathcal{O}_c \Phi$, where $\mathcal{O}_c$ is an operator in the CFT of dimension $\Delta$.  If $\Delta < 2$, then this operator is relevant, and we expect the theory to flow to a fixed point where $\Phi$ has dimension $1+(2-\Delta) \equiv 1 + \gamma$.  

\vskip 4pt

$\boldsymbol{\mu > m}$:
We will first consider the case where inflation takes place at scales for which $\Phi$ is still coupled to the CFT.  The anomalous dimension for $\Phi$ arises through wavefunction renormalization, and can be treated by modifying the K\"ahler potential $K = Z(\mu/ \Lambda)\Phi^{\dag} \Phi$. 
As a result, the leading term in the potential energy is now $V \sim |\sigma|^4 Z^{-1}(\mu / \Lambda)$.  During inflation, $\Phi$ acquires a vev that breaks conformal invariance.  Schematically, we may think of this as the scale where the CFT is cut off and thus we write $\mu \sim y \Phi$.  The effective potential for $\Phi$ now takes the form $V \sim \sigma^4 (y \Phi / \Lambda)^{2\gamma}$.  It is then easy to see that $\varepsilon \sim 2\gamma \frac{M_{\rm pl}^2}{\Phi^2}$ and thus the anomalous dimensions must be extremely small in order to have slow-roll (assuming $\Phi < M_{\rm pl}$).  Obviously, such tiny dimensions will not be sufficient to solve the eta problem arising from higher-dimension operators in the K\"ahler potential.

\vskip 4pt

$\boldsymbol{\mu < m}$:
We could instead try to 
break conformal invariance at a scale that is higher than the scale of inflation.  The low-energy theory is then the same as the theory without the coupling to the CFT, but the coefficients $c_1$ and $c_2$ may be modified by the running at higher energies.  We would like these coefficients to be small, and thus they should run to zero.  
In the model in \S\ref{sec:susy}, the dangerous couplings were of the form $\Phi^2 X^\dag X$, where $\Phi$ was coupled to the CFT but $X$ was not.  Because $X$ was not coupled to the CFT, there was no way to generate this operator through CFT couplings alone, and thus it had to run to zero at the fixed point.  Furthermore, only $U(1)$ breaking operators could contribute to the potential, and so all corrections of the form $(\Phi^{\dag} \Phi)^n$ could be ignored.

Unfortunately, this is not true of the linear model because here the dangerous operators include products of $\Phi$ like $(\Phi^{\dag} \Phi)^2$.  We should therefore expect this operator to be produced by diagrams involving the CFT alone.  Indeed, if $\mathcal{O}_c = Q \tilde{Q}$, as before, it should be clear that we will generate dangerous dimension-six operators through loops of $Q$ and $\tilde{Q}$.  After breaking conformal invariance at a scale $m$, the effective theory below that scale will therefore include K\"ahler potential terms suppressed only by powers of $m$---{\it e.g.},~$K \supset \frac{c}{m^2} (\Phi^{\dag} \Phi)^2$.  Even though the coefficients will be set by the coupling $y$, the contributions to $\varepsilon$ and $\eta$ will be enhanced by factors of $\frac{M_{\rm pl}^2}{m^2}$.  This will make it impossible to suppress these corrections through RG flow.

\subsection{Lessons Learned}
\label{sec:lessons}

There are two basic lessons that one should take away from this example:

\begin{itemize}
\item[-] Modifying the dimension of the inflaton effectively introduces an overall power of $\phi^{n \gamma}$ to the potential energy.  For small-field inflation, slow-roll requires a potential of the form $V(\phi) \sim V_0 + \delta V(\phi)$.  The overall power of $\phi^{n \gamma}$ from wavefunction renormalization will often appear in front of $V_0$, ruining the flatness of the potential.  This typically requires us to break conformal invariance above the scale of inflation. Then, wavefunction renormalization only renormalizes the couplings, but does not introduce overall factors of the field $\phi$.

\item[-] The second problem is that coupling the inflaton to other sectors may reintroduce the very operators one is looking to suppress.  One should expect this to occur any time the dangerous terms in the potential involve only fields that are coupled to a CFT.  
%This is generically a problem. 
Fortunately, when the inflationary potential is protected by an approximate symmetry, CFT couplings that respect this symmetry will not reintroduce the problem (see \S\ref{sec:nonR}).  This approximate symmetry reduces the number of dangerous corrections to those that break the symmetry, making it easier to suppress all the contributions to eta.
\end{itemize}

\subsection{Comments on Large-Field Models}
\label{sec:large}

The UV-sensitivity of inflation is significantly enhanced in large-field models---{\it i.e.}~in models in which the inflaton field traverses a super-Planckian distance $\Delta \phi \gg M_{\rm pl}$. In this case, an infinite series of corrections has to be considered, ${\cal O}_4 \sum_n \frac{\phi^n}{M_{\rm pl}^n}$. It is tempting to think that wavefunction renormalization could systematically suppress the higher-order terms and therefore control the whole series of terms at once.
To highlight some of the challenges we see with this idea, we consider the supergravity model of chaotic inflation of Kawasaki, Yamaguchi and Yanagida~\cite{Kawasaki:2000yn}.
Using a shift-symmetric K\"ahler potential, $K=(\Phi + \Phi^\dagger)^2 + X X^\dagger$, and a linear superpotential, $W=m X \Phi$, it was shown that the F-term potential is $V \approx m^2 \varphi^2$, where $\varphi \equiv {\rm Im}(\Phi) \gtrsim M_{\rm pl}$ \cite{Kawasaki:2000yn}.

Despite appearance, there are several obstacles to using sequestering in a large-field model of this type.  First of all, when the field range is larger than $M_{\rm pl}$, it is not clear what it means to have a single effective description.  Because the UV cutoff is at most $M_{\rm pl}$, over the course of inflation, the masses of the light fields can become larger than the cutoff, while similarly, heavy fields may become light.  Even if we were able to avoid this conceptual problem, it still seems unlikely that we could construct a CFT with the right properties.  As we have just seen, it is not enough to simply couple the inflaton to a CFT; one must couple the inflaton to a CFT without breaking the global symmetry that protects the potential.  In the case of a shift symmetry, this seems to be a significant challenge.  Because the shift symmetry involves the imaginary part of $\Phi$, we will not be able to do this through a superpotential coupling without reintroducing the dangerous terms in the K\"ahler potential.  One could gauge the shift symmetry, but this typically requires explicit breaking of conformal invariance and gives mass to the gauge boson.
These difficulties with large-field models further support the intuition that a UV-completion is necessary when the field range is on the order of $M_{\rm pl}$ \cite{Eva, Liam}.  

\newpage
\section{Non-SUSY Examples}
\label{sec:nonSUSY}

So far we have focused on supersymmetric examples because they offer simple means of creating radiatively stable and technically natural models.  However, non-supersymmetric examples with the same features also exist \cite{ArkaniHamed:2003mz,Kaplan:2003aj,Sundrum:2009ii} and 
in this section we apply our ideas to these cases.

We are then obliged to consider non-minimal couplings to gravity of the schematic form $\xi \phi^2 R$ (\S\ref{sec:nonMin}). A concern in using RG flow to suppress these couplings is that the couplings $\xi$ may not flow to zero but to a finite value at the fixed point.
Indeed, in many cases $\xi = \frac{1}{6}$ at the fixed point of the RG flow (\S\ref{sec:RGFlow}). We will discuss under what conditions this problem is absent (\S\ref{sec:nonS}) and connect this to the conditions for successful sequestering in our supersymmetric models.

\subsection{Non-Minimal Couplings to Gravity}
\label{sec:nonMin}

Consider the action for a complex scalar field
\beq \label{classical}
S = \int d^4 x \sqrt{-g}\, \bigl[ \partial_\mu \phi \partial^\mu \phi^{\dag} - V(\phi, \phi^{\dag}) \bigr]\, ,
\eeq
where $V(\phi, \phi^{\dag})$ is assumed to be radiatively stable and technically natural.  
We will be agnostic about the precise mechanism by which we create a radiatively stable potential.  
In the absence of other contributions to the potential, this model is assumed to inflate.
To this action we add non-minimal couplings to gravity of the form
\beq
\label{equ:nonmini}
\delta S  = - \int d^4 x \sqrt{-g}\, R \bigl[c_1 M_{\rm pl} (\phi + \phi^{\dag}) + c_2 (\phi^2 + \phi^{\dag 2}) + c_3 \phi^{\dag} \phi \bigr] + \cdots\ ,
\eeq
where $R$ is the Ricci scalar.  During inflation $R \sim \frac{V}{M_{\rm pl}^2}$ and these terms can give order one corrections to $\eta$.

Naively, it might seem that if $\phi$ were to acquire an anomalous dimension, these terms would run to zero.  However, as we saw for the model in \S\ref{sec:linear}, we must check that $c_{i} \neq 0$ are not fixed points of the RG flow.  One important difference between the PNGB model in \S\ref{sec:susy} and the linear model in \S\ref{sec:linear} was that the dangerous terms in the PNGB model involved fields that did not couple to the CFT and thus could not be fixed points of the RG.  For the couplings in (\ref{equ:nonmini}), because of the universal coupling of gravity, the curvature terms can be generated by the CFT and thus we should expect some of these terms to be non-zero at the fixed point.

In this section we will discuss the RG flow of the curvature couplings in (\ref{equ:nonmini}).
Our main conclusion will be that $c_3$ necessarily runs to a fixed point, 
%(when kinetic term in given by the canonical form in (\ref{classical})), 
while $c_1$ and $c_2$ may or may not run to zero (this will ultimately depend on details of the model).  The intuitive reason for this result is that $c_3$ is directly tied to the kinetic term in (\ref{classical}).  For example, for the action of a free field to be conformally invariant in a curved background, $c_3 = \frac{1}{6}$.  This curvature term is equivalent to the dimension-six operators in supergravity proportional to the canonical K\"ahler potential. 

The other terms in (\ref{equ:nonmini}) may or may not be generated depending on how the inflaton is coupled to the CFT.  For example, the action may have an approximate $U(1)$ symmetry where $\phi \to e^{i\alpha} \phi$.  Just like in our SUSY example, if the couplings to the CFT are such that $U(1)$ breaking couplings cannot be generated, then $c_{1}$ and $c_2$ will run to zero under RG.

\subsection{Renormalization Group Flow}
\label{sec:RGFlow}

We digress briefly to illustrate in a toy example how interactions can drive the couplings to finite values.
This should be understood as an analogy for the problems one generically faces when coupling (\ref{classical}) and (\ref{equ:nonmini}) to a CFT.
For simplicity, we restrict  to the example of a {\it real} free scalar~$\phi$. 
We write down the most general coordinate invariant action with couplings that may depend on the RG scale, but that are independent of the background
\beq
\label{equ:L}
\frac{ \mathcal{L}}{\sqrt{-g}} = M_{\rm pl}^2 R + a_1 R^2 + a_2 R_{\mu \nu} R^{\mu \nu} + Z \partial_{\mu} \phi \partial^{\mu}\phi + a_3 M_{\rm pl} \phi R - \frac{\xi_1}{2} \phi^2 R + \cdots
\eeq
We then couple the inflaton $\phi$ to another field $\psi$ through a $\lambda \phi^2 \psi^2$ interaction. 
Here, the field $\psi$ should be viewed as the analogue of the CFT flavors $Q$ in the previous sections. We furthermore couple $\psi$ to gravity in a non-minimal way, $\xi_2 \psi^2 R$.

There are several approaches to studying the RG flow of the curvature couplings in (\ref{equ:L}) ({\it e.g.}~Refs.~\cite{'tHooft:1974bx,Birrell:1982ix,Buchbinder:1992rb}).  
The approach taken by 't Hooft and Veltman in \cite{'tHooft:1974bx} is the most direct, as it computes the divergences directly.  When we don't include graviton loops, their analysis simplifies dramatically.  Because the couplings are independent of the background, one is free to choose a simple background in which calculations can be done easily.  Specifically, we choose the metric to be conformally Minkowski, $g_{\mu \nu} = F(x) \eta_{\mu \nu}$.  
The relevant terms of the action then are
\beq
S = \int d^4 x\, \Bigl[ \frac{1}{2} F \partial_{\mu} \phi \partial^{\mu} \phi+ \frac{1}{2} F \partial_{\mu} \psi \partial^{\mu} \psi  - \frac{\lambda}{4} F^2 \phi^2 \psi^2 \Bigr]    - \frac{\xi_1}{2} \phi^2F R  - \frac{\xi_2}{2} \psi^2 F R + \cdots\  ,
\eeq
where the indices are contracted with the flat metric $\eta_{\mu \nu}$.  In terms of the canonically-normalized field $\phi \to \phi \sqrt{F}^{-1}$ and similarly for $\psi$, the action becomes
\begin{align}
S &= \int d^4x\ \Bigl[ \frac{1}{2} \partial_{\mu} \phi \partial^{\mu} \phi + \frac{1}{2} \phi F^{-1} \partial_\mu F \partial^{\mu} \phi - \frac{1}{8} \phi^2 F^{-2} \partial_\mu F \partial^{\mu} F +\ \{ \phi \to \psi \} \ - \frac{\lambda}{4} \phi^2 \psi^2 \Bigr] \\
 &\hspace{3cm}- \frac{\xi_1}{2} \phi^2 R  - \frac{\xi_2}{2} \psi^2 R + \dots \nonumber
\end{align}
The beta functions can now be determined using field theory in flat space with these additional couplings and  treating $X_{\mu} \equiv F^{-1} \partial_{\mu} F$ as an external field.  At one loop, there are four logarithmically divergent diagrams that sum to 
\vspace{0.1cm}
\beq
 \includegraphicsbox[scale=.4]{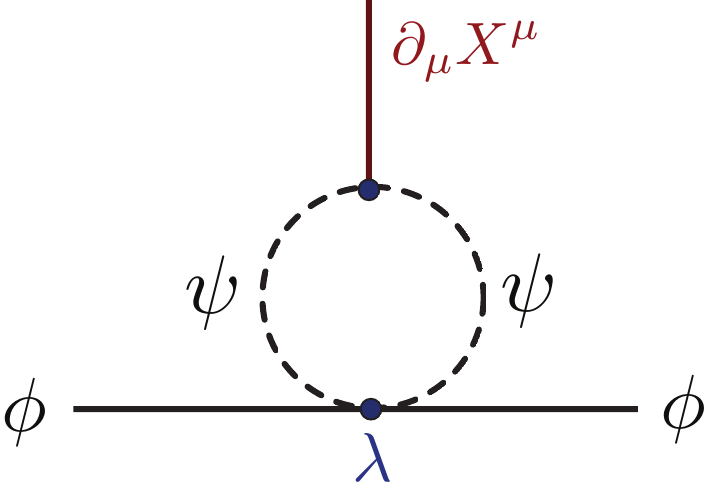} \ \ + \ \  \includegraphicsbox[scale=.4]{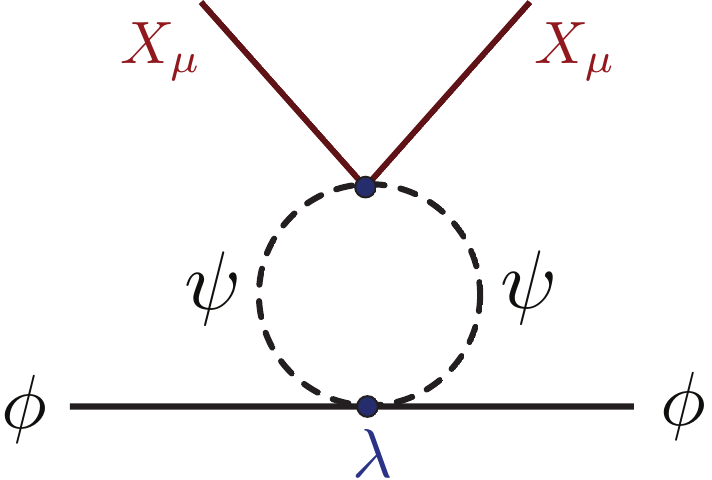} \ \ + \ \ \includegraphicsbox[scale=.4]{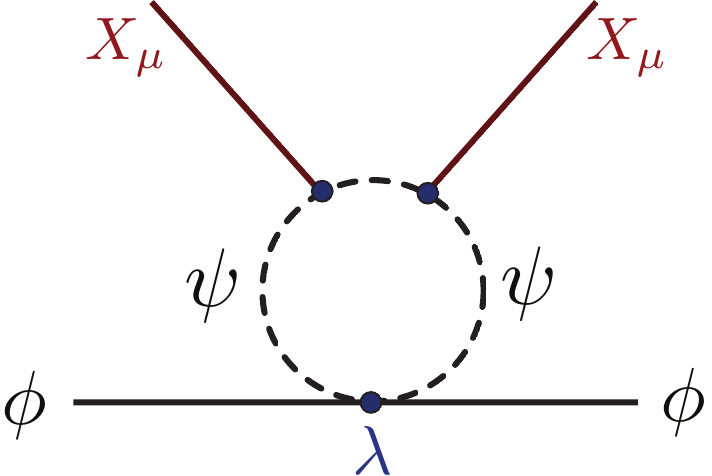} \ \ + \ \ \includegraphicsbox[scale=.4]{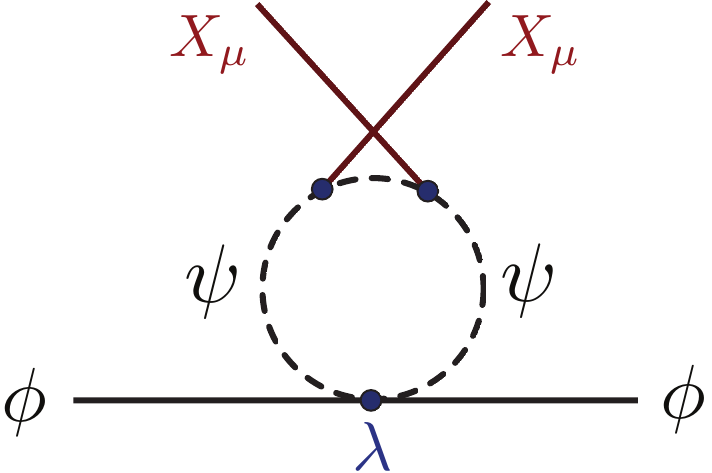}  
 \eeq
\begin{align}
&= \ \ - \frac{i \lambda}{16 \pi^2} \log\Bigl(\frac{\mu}{\Lambda} \Bigr) F^{-1} \Bigl(\frac{1}{8} X_{\mu} X^{\mu} + \frac{1}{4} \partial_{\mu} X^{\mu} \Bigr) \phi^2 \\
&= \ \ - \frac{1}{12}  \frac{i \lambda}{16 \pi^2} \log\Bigl(\frac{\mu}{\Lambda} \Bigr)  \phi^2 {R}\, , \label{term1}
\end{align}
where we have used
\beq
{R} = 3 F^{-2} \partial_{\mu}\partial^{\mu} F - \frac{3}{2} F^{-3}  \partial_{\mu} F \partial^{\mu} F = 3 F^{-1} \Bigl(\partial_\mu X^\mu + \frac{1}{2} X_\mu X^\mu \Bigr)\, .
\eeq
Eqn.~(\ref{term1}) hence gives a logarithmically divergent contribution to $\xi_1$.  By exchange symmetry, there is of course an identical contribution with $\phi \to \psi$.  Similarly, for non-zero $\xi_2$ there is a contribution of the form
\beq
\includegraphicsbox[scale=.45]{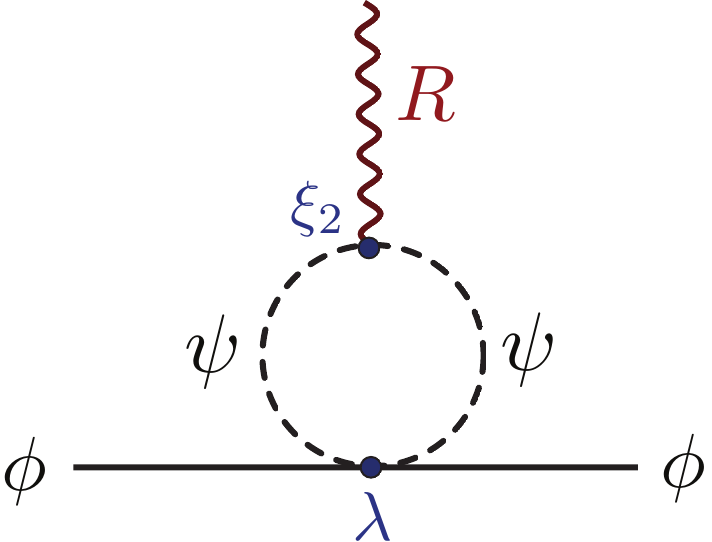} 
\ \ = \ \ + \frac{i \lambda}{32 \pi^2} \log\Bigl(\frac{\mu}{\Lambda} \Bigr) \xi_2 \phi^2 {R}\, .
\label{equ:xi2}
\eeq
Putting these results together, we find that the one-loop beta functions for the curvature couplings $\xi_1$ and $\xi_2$ are
\bea
\beta_{\xi_1} \equiv \frac{\partial \xi_1}{\partial \log(\mu)} = \frac{( \xi_2 - \frac{1}{6})}{16 \pi^2} \frac{\lambda}{2}\, , \\
\beta_{\xi_2} \equiv \frac{\partial \xi_2}{\partial \log(\mu)} = \frac{(\xi_1 - \frac{1}{6})}{16 \pi^2} \frac{\lambda}{2}\, .
\eea
As a result, we see that $\xi_{1}$ and $\xi_2$ will run to the conformal coupling of $\xi_{1}=\xi_2 = \frac{1}{6}$.\footnote{If we had included graviton loops we would have found that  $\xi_{1}$ and $\xi_2$ are driven to $\frac{1}{6}$ even when $\lambda \to 0$ \cite{'tHooft:1974bx} (this is the conformal fixed point of a free scalar). }

This important result, in fact, extends beyond our toy example.
As discussed in \cite{Buchbinder:1992rb}, it can be shown that this structure of the one-loop beta functions is generic.  In general, one can write the counterterms in the form $Z_{\xi_1} = Z_2 \xi_1 + Z_3$ (and similarly for $\xi_2$).  Not surprisingly, one can then show that $Z_2 = Z_m$, the renormalization of the mass term.  At one loop, it can furthermore be shown that $\xi_{1} = \xi_2 = \frac{1}{6}$ is a fixed point, and thus we have $Z_3 = -\frac{1}{6} Z_2$~\cite{Buchbinder:1992rb}.

\subsection{Conformal Sequestering}
\label{sec:nonS}

Let us discuss the implications of the above considerations for sequestering in non-supersymmetric models.  
Going back to the complex scalar in (\ref{classical}) and (\ref{equ:nonmini}), we conclude that:
\begin{enumerate}
\item[i)]  If the inflaton is a scalar $\phi$ with canonical kinetic term, then any type of running will cause $c_3$ to run to a non-zero fixed point.  This should not be surprising, since conformal invariance of the kinetic term requires that $c_3 = \frac{1}{6}$ and thus we should expect the CFT to drive it to that value.  In that sense, we may think of the term proportional to $c_3$ as being on par with the kinetic terms.  In SUSY, this role is played by the $e^{K / M_{\rm pl}^2}$ prefactor in the scalar potential.

\item[ii)] The second dangerous coupling $c_2$ may or may not run to zero depending on the precise coupling to the CFT.  This can be seen from the toy model of \S\ref{sec:RGFlow} by thinking of the complex scalar as two real fields ($\phi = \phi_1 + i\phi_2$).  By adding the coupling $\psi^2(\phi_1^2 - \phi_2^2)$, a non-zero beta function for $c_2$ is generated.
\end{enumerate}
To resolve the problem of conformal couplings, we use non-canonical kinetic terms, just like in the SUSY model of \S\ref{sec:susy}.  As before, we could have constructed a model such that the inflaton arises as a 
PNGB~\cite{ArkaniHamed:2003mz,Kaplan:2003aj}.  
Specifically, 
we will now assume that the model in (\ref{classical}) has a softly broken $U(1)$ symmetry under which $\phi \to e^{i \alpha} \phi$.  In the low-energy theory, the potential should lead to the vev $\phi = (f + \rho(x))e^{i \varphi(x)}$, where $\varphi(x)$ will be the inflaton.  If we couple $\phi$ to a CFT in such a way that the $U(1)$ is unbroken, then the CFT will not reintroduce dangerous terms into the potential.  Specifically, let us add to the potential the dangerous curvature couplings,
\beq
S = \int d^4 x \sqrt{-g}\, \Bigl[ \partial_\mu \phi \partial^\mu \phi^{\dag} - V(\phi, \phi^{\dag}) - R \bigl[c_1 M_{\rm pl} (\phi + \phi^{\dag}) + c_2 (\phi^2 + \phi^{\dag 2}) + c_3 \phi^{\dag} \phi \bigr]\Bigr] \, .
\eeq
As in our SUSY model, it will be more convenient to forbid the $c_1$ term by a $\mathbb{Z}_2$ symmetry, rather than requiring large anomalous dimensions.  We expect that $c_3 \simeq \frac{1}{6}$ after RG flow, but it should be clear that this term does not contribute to the potential for $\varphi$.  As a result, the only terms of interest are the terms proportional to $c_2$.  However, because these terms break the $U(1)$~symmetry, they must run to zero at the fixed point as long as the couplings to the CFT are $U(1)$~invariant.
RG flow can therefore indeed protect the potential of the PNGB $\varphi$.

In our SUSY models, the analogue of the coupling $c_2$ are terms in the K\"ahler potential of the form $\Phi^2$.  In field theory, these terms vanish (since $\int d^4 \theta\, \Phi^2 =0$ because $\Phi$ is chiral), but in supergravity they contribute to the scalar potential.  As we discussed above, these terms break the $U(1)$ symmetry and cannot be generated in the CFT.  As a result, they run to zero with the same anomalous dimensions as the other dimension-six operators.

\section{Comparison with Gauged Models}
\label{sec:gauged}

The mechanism described in this paper is a low-energy solution to the eta problem, in the sense that the corrections to $\eta$ are of order one in the UV, but their small size in the IR is understood by field theory dynamics alone.  
The addition of order one symmetry breaking operators at the Planck scale was inspired by the well-known lore that nonperturbative quantum gravity effects break all {\it global} symmetries~\cite{Kallosh:1995hi,Kamionkowski:1992mf, Holman:1992us}.
We now briefly contrast this solution to the eta problem with the possibility that the symmetry protecting the inflaton potential arises from a {\it gauge} symmetry. In that case the standard black hole evaporation arguments do not apply and the symmetry breaking operators may be suppressed in the UV.
This can also be considered a low-energy solution to the eta problem, but as we now explain there are important differences between the two ideas.

\vskip 4pt
For concreteness, consider the case where the approximate shift symmetry is associated with the Goldstone boson of a spontaneously broken global $U(1)$ symmetry.  
Assume that the global $U(1)$ becomes a gauge symmetry in the UV. 
If the $U(1)$ has anomalies which are cancelled by an axionic shift symmetry, the gauge symmetry is broken spontaneously.  Nevertheless, the breaking is such that the $U(1)$ symmetry survives in the IR as an approximate global symmetry that is only broken by nonperturbative effects.  Ultimately, one requires these nonperturbative effects to give rise to the small slope of the inflaton potential.
Once the symmetry is gauged, any operator in the action is required to be gauge-invariant.  Nonperturbative effects are gauge-invariant because of the shift of the theta angle under the symmetry.  As a result, any operator that contributes to the potential has a coefficient that is exponentially small.  This is also consistent with the black hole argument because there are long range forces associated with the charge.  

A downside of gauging the shift symmetry is that it demands that the UV-completion provides the right nonperturbative effects to the inflaton potential, but not others.  These effects do not arise in the pure U(1) gauge theory, but can arise from effects in the UV-completion.  For example, in string theory, these nonperturbative effects are generated by stringy instantons~\cite{Blumenhagen:2009qh}, while in extra-dimensional scenarios \cite{ArkaniHamed:2003mz, Kaplan:2003aj}, they can arise from non-local potentials for higher-dimensional gauge fields \cite{Hosotani:1983xw}.  One could also try to generate the potential using field theory instantons by coupling to a non-abelian gauge field, but to our knowledge this has not been done.
In the absence of computing the nonperturbative contributions to the potential in the effective theory, the gauged models hence do make some assumption about the structure of the UV-completion. However, arguably this assumption is much weaker and much more plausible than assuming the absence of dangerous symmetry breaking operators in models with global symmetries.

Furthermore, the gauged $U(1)$ symmetry should be contrasted with the gauged $\mathbb{Z}_2$ symmetry that we required in our models.  The $\mathbb{Z}_2$ symmetry is exact, and we do not need any further knowledge of its origin in the UV-completion.  The question is only whether UV-completions allow for exact $\mathbb{Z}_N$ symmetries, to which string theory's answer is yes.  If our models had required this symmetry to be broken in some mild way, we would also have needed to know more about the UV origin of the discrete symmetry. However, in the examples we presented this was not the case.

Finally, one might wonder if our mechanism is dual to the gauged models in the sense of AdS/CFT.  Specifically, one could construct a warped geometry where the inflaton arises from a five-dimensional gauge field like in \cite{ArkaniHamed:2003mz, Kaplan:2003aj}.  In the extra-dimensional formulation, the suppression of the higher-dimension operators arises from the size of the AdS region.  In the CFT description, the suppression is the result of strong dynamics.  However, our models are {\it weakly-coupled} CFTs and so the dual description is in terms of a strongly-coupled model of gravity.  While the ideas are closely related, it is important that our models are not just a dual description of weakly-coupled string or extra-dimensional constructions.

%\newpage
\section{Conclusions}
\label{sec:conclusion}

It is rare that low-energy physics depends sensitively on Planck-suppressed contributions.
Inflation is one of the few examples where understanding these corrections to the action is absolutely essential.
It is important to realize that the eta problem is independent of the energy scale of inflation and is equally severe for high-scale and low-scale models. 
In this paper we have presented a new solution to this problem.

While most solutions to the eta problem assume low-energy symmetries and the absence of symmetry breaking operators in the UV, we have shown that appropriate couplings of the inflaton to a conformal sector allow control over these corrections in effective field theory.
This has allowed us to relax some of the commonly made assumptions about the UV structure of the theory.
We have presented explicit examples to illustrate how conformal sequestering can suppress the effects of shift symmetry violating terms in the UV. The low-energy theory then remains approximately shift symmetric and has a small eta parameter even though the shift symmetry is badly broken in the UV.

We summarize what we have learned as a guide for future applications of our idea. %model-building.
The theory has to contain the following elements to allow a successful decoupling of higher-dimension corrections to the inflationary action:

\begin{enumerate}
\item {\it Symmetries of the renormalizable action}

In the basis where the inflaton has canonical kinetic term, the potential may be split into a renormalizable part, $V_0(\phi)$, and non-renormalizable corrections, $\delta V(\phi)$,
\beq
V(\phi) = V_0(\phi) + \delta V(\phi)\, . \nonumber
\eeq
Technical naturalness and radiative stability require that the renormalizable action has certain symmetries. Different models will achieve these desirable features in different ways.
In this paper we used a combination of supersymmetry and an approximate shift symmetry to protect the renormalizable part of the potential.

\item {\it Symmetries of the coupling to the CFT}

To prevent generating dangerous operators via the coupling to the CFT itself, we require that the couplings to the conformal sector respect the same symmetries as the renormalizable action.
This ensures that the couplings of the dangerous operators flow to zero and not to some finite value at the fixed point.

\item {\it UV corrections}

In the non-renormalizable part of the potential, $\delta V(\phi)$, we allow arbitrary breaking of the symmetries of the renormalizable part of the potential.
RG flow will suppress the couplings of these higher-dimension operators, so that the full action in the IR has the (accidental) symmetries of the renormalizable potential.

\end{enumerate}

We have shown in a variety of examples that these requirements can be fullfilled in a technically natural way.
In our most explicit example, in \S\ref{sec:susy}, we computed the anomalous dimension of the inflaton---exactly via $a$-maximization and at one loop---and showed that our mechanism involves only weakly-coupled physics. 
%We therefore have confidence in the computational control. 
Going to stronger coupling is likely to increase the efficiency of sequestering, but reduces the control over the field theory computations. It might be interesting to study this regime in the gravity dual \cite{AdSCFT}.  

\vskip 4pt
% However, this is not to say that 
 Finally, we would like to be clear that our work is not meant to be read as claiming that understanding the UV-completion of inflationary models
 % small-field models 
 is not important ({\it cf.}~\S\ref{sec:comments}).  Our goal has been to explore the possibility of solving the eta problem while being agnostic about the effects of a UV-completion.
 In small-field models, we believe that this is possible through RG flow in the low-energy effective theory, provided approximate continuous  symmetries in the IR and a discrete symmetry in the UV.  However, even in this context, it would be very useful to understand the origin of approximate symmetries within a UV-complete framework.  It may be the case that engineering these approximate symmetries requires special features that ultimately suppress dangerous Planck-suppressed contributions to the potential.  Nevertheless, the K\"ahler corrections in our models are controlled in field theory and one only must only explain the origin of the superpotential.  For this reason, one could hope to build a model in string theory using only topological information (see, for example, \cite{Blumenhagen:2005mu, Douglas:2006es, Blumenhagen:2006ci, Denef:2008wq}). 
 Furthermore, we have exhibited classes of inflationary models for which UV corrections cannot be decoupled by RG flow. For these models understanding the UV-completion is essential.

\subsubsection*{Acknowledgements}

We are grateful to Nima Arkani-Hamed, Nathaniel Craig, Anatoly Dymarsky, Jonathan Heckman, Liam McAllister, Michele Papucci, Soo-Jong Rey, Leonardo Senatore, Eva Silverstein, Matt Sudano, Tomer Volansky, and Brian Wecht for discussions.
We thank Liam McAllister for extremely helpful comments on a draft.
D.B.~wishes to express special thanks to Anatoly Dymarsky, Shamit Kachru, Igor Klebanov and Liam McAllister for collaboration on related questions.
D.B.~thanks the Mitchell Institute for Fundamental Physics and Astronomy at Texas A\&M for hospitality and the opportunity to present this work.
D.G.~thanks the Kavli Institute for Theoretical Physics for hospitality while this work was completed.
The research of D.B.~is supported by the National Science Foundation under PHY-0855425, AST-0506556 and AST-0907969. The research of D.G.~is supported by the Department of Energy under grant number DE-FG02-90ER40542.

 \newpage
%\section*{References}

\vfil

\begin{thebibliography}{10}

%\cite{McAllister:2007bg}
\bibitem{McAllister:2007bg}
  L.~McAllister and E.~Silverstein,
  ``String Cosmology: A Review,''
  Gen.\ Rel.\ Grav.\  {\bf 40}, 565 (2008)
  [arXiv:0710.2951 [hep-th]].
  %%CITATION = GRGVA,40,565;%%

%\cite{Burgess:2007pz}
\bibitem{Burgess:2007pz}
  C.~P.~Burgess,
  ``Lectures on Cosmic Inflation and its Potential Stringy Realizations,''
  PoS {\bf P2GC}, 008 (2006)
  [Class.\ Quant.\ Grav.\  {\bf 24}, S795]
  [arXiv:0708.2865 [hep-th]].
  %%CITATION = POSCI,CARGESE2007,003;%%

%\cite{Baumann:2008aq}
\bibitem{Baumann:2008aq}
  D.~Baumann {\it et al.}  [CMBPol Study Team Collaboration],
  ``CMBPol Mission Concept Study: Probing Inflation with CMB Polarization,''
  AIP Conf.\ Proc.\  {\bf 1141}, 10 (2009)
  [arXiv:0811.3919 [astro-ph]].
  %%CITATION = APCPC,1141,10;%%

%\cite{Baumann:2009ni}
\bibitem{Baumann:2009ni}
  D.~Baumann and L.~McAllister,
  ``Advances in Inflation in String Theory,''
  Ann.\ Rev.\ Nucl.\ Part.\ Sci.\  {\bf 59}, 67 (2009)
  [arXiv:0901.0265 [hep-th]].
  %%CITATION = ARNUA,59,67;%%
  
   %\cite{Stewart:1996ey}
\bibitem{Stewart:1996ey}
  E.~D.~Stewart,
  ``Flattening the Inflaton's Potential with Quantum Corrections,''
  Phys.\ Lett.\  B {\bf 391}, 34 (1997)
  [arXiv:hep-ph/9606241].
  %%CITATION = PHLTA,B391,34;%%
  
  %\cite{Stewart:1997wg}
\bibitem{Stewart:1997wg}
  E.~D.~Stewart,
  ``Flattening the Inflaton's Potential with Quantum Corrections. II,''
  Phys.\ Rev.\  D {\bf 56}, 2019 (1997)
  [arXiv:hep-ph/9703232].
  %%CITATION = PHRVA,D56,2019;%%
 
 %\cite{Lyth:1998xn}
\bibitem{Lyth:1998xn}
  D.~H.~Lyth and A.~Riotto,
  ``Particle Physics Models of Inflation and the Cosmological Density
  Perturbation,''
  Phys.\ Rept.\  {\bf 314}, 1 (1999)
  [arXiv:hep-ph/9807278].
  %%CITATION = PRPLC,314,1;%%
  
  %\cite{Freese:1990rb}
\bibitem{Freese:1990rb}
  K.~Freese, J.~A.~Frieman and A.~V.~Olinto,
  ``Natural Inflation with Pseudo-Nambu-Goldstone Bosons,''
  Phys.\ Rev.\ Lett.\  {\bf 65}, 3233 (1990).
  %%CITATION = PRLTA,65,3233;%%
  
  %\cite{Kallosh:1995hi}
\bibitem{Kallosh:1995hi}
  R.~Kallosh, A.~D.~Linde, D.~A.~Linde and L.~Susskind,
  ``Gravity and Global Symmetries,''
  Phys.\ Rev.\  D {\bf 52}, 912 (1995)
  [arXiv:hep-th/9502069].
  %%CITATION = PHRVA,D52,912;%%
  
  %\cite{Kamionkowski:1992mf}
\bibitem{Kamionkowski:1992mf}
  M.~Kamionkowski and J.~March-Russell,
  ``Planck Scale Physics and the Peccei-Quinn Mechanism,''
  Phys.\ Lett.\  B {\bf 282}, 137 (1992)
  [arXiv:hep-th/9202003].
  %%CITATION = PHLTA,B282,137;%%
  
%\cite{Holman:1992us}
\bibitem{Holman:1992us}
  R.~Holman, S.~D.~H.~Hsu, T.~W.~Kephart, E.~W.~Kolb, R.~Watkins and L.~M.~Widrow,
  ``Solutions to the Strong CP Problem in a World with Gravity,''
  Phys.\ Lett.\  B {\bf 282}, 132 (1992)
  [arXiv:hep-ph/9203206].
  %%CITATION = PHLTA,B282,132;%%
  
%\cite{Kamionkowski:1992ax}
\bibitem{Kamionkowski:1992ax}
  M.~Kamionkowski and J.~March-Russell,
  ``Are Textures Natural?,''
  Phys.\ Rev.\ Lett.\  {\bf 69}, 1485 (1992)
  [arXiv:hep-th/9201063].
  %%CITATION = PRLTA,69,1485;%%
  
  %\cite{Silverstein:2008sg}
\bibitem{Eva}
  E.~Silverstein and A.~Westphal,
  ``Monodromy in the CMB: Gravity Waves and String Inflation,''
  Phys.\ Rev.\  D {\bf 78}, 106003 (2008)
  [arXiv:0803.3085 [hep-th]].
  %%CITATION = PHRVA,D78,106003;%%
  
  %\cite{McAllister:2008hb}
\bibitem{Liam}
  L.~McAllister, E.~Silverstein and A.~Westphal,
  ``Gravity Waves and Linear Inflation from Axion Monodromy,''
  arXiv:0808.0706 [hep-th].
  %%CITATION = ARXIV:0808.0706;%%
  
  %\cite{Berg:2009tg}
\bibitem{Berg:2009tg}
  M.~Berg, E.~Pajer and S.~Sjors,
  ``Dante's Inferno,''
  arXiv:0912.1341 [hep-th].
  %%CITATION = ARXIV:0912.1341;%%

  
  %\cite{ArkaniHamed:2003mz}
\bibitem{ArkaniHamed:2003mz}
  N.~Arkani-Hamed, H.~C.~Cheng, P.~Creminelli and L.~Randall,
  ``Pseudonatural Inflation,''
  JCAP {\bf 0307}, 003 (2003)
  [arXiv:hep-th/0302034].
  %%CITATION = JCAPA,0307,003;%%
  
  %\cite{Kaplan:2003aj}
\bibitem{Kaplan:2003aj}
  D.~E.~Kaplan and N.~J.~Weiner,
  ``Little Inflatons and Gauge Inflation,''
  JCAP {\bf 0402}, 005 (2004)
  [arXiv:hep-ph/0302014].
  %%CITATION = JCAPA,0402,005;%%
  
  
  %\cite{Susskind:1978ms}
\bibitem{Susskind:1978ms}
  L.~Susskind,
  ``Dynamics of Spontaneous Symmetry Breaking in the Weinberg-Salam Theory,''
  Phys.\ Rev.\  D {\bf 20}, 2619 (1979).
  %%CITATION = PHRVA,D20,2619;%%
  
  %\cite{Weinberg:1975gm}
\bibitem{Weinberg:1975gm}
  S.~Weinberg,
  ``Implications of Dynamical Symmetry Breaking,''
  Phys.\ Rev.\  D {\bf 13}, 974 (1976).
  %%CITATION = PHRVA,D13,974;%%
  
 %\cite{Witten:1981nf}
\bibitem{Witten:1981nf}
  E.~Witten,
  ``Dynamical Breaking of Supersymmetry,''
  Nucl.\ Phys.\  B {\bf 188}, 513 (1981).
  %%CITATION = NUPHA,B188,513;%%
  
     %\cite{Nelson:2000sn}
\bibitem{Nelson:2000sn}
  A.~E.~Nelson and M.~J.~Strassler,
  ``Suppressing Flavor Anarchy,''
  JHEP {\bf 0009}, 030 (2000)
  [arXiv:hep-ph/0006251].
  %%CITATION = JHEPA,0009,030;%%

  %\cite{Luty:2001jh}
\bibitem{Luty:2001jh}
  M.~A.~Luty and R.~Sundrum,
  ``Supersymmetry Breaking and Composite Extra Dimensions,''
  Phys.\ Rev.\  D {\bf 65}, 066004 (2002)
  [arXiv:hep-th/0105137].
  %%CITATION = PHRVA,D65,066004;%%

%\cite{Luty:2001zv}
\bibitem{Luty:2001zv}
  M.~Luty and R.~Sundrum,
  ``Anomaly-mediated Supersymmetry Breaking in Four Dimensions, Naturally,''
  Phys.\ Rev.\  D {\bf 67}, 045007 (2003)
  [arXiv:hep-th/0111231].
  %%CITATION = PHRVA,D67,045007;%%
 
 
%\cite{Binetruy:1996xj}
\bibitem{Binetruy:1996xj}
  P.~Binetruy and G.~R.~Dvali,
  ``D-term Inflation,''
  Phys.\ Lett.\  B {\bf 388}, 241 (1996)
  [arXiv:hep-ph/9606342];
  
  E.~Halyo,
  ``Hybrid Inflation from Supergravity D-terms,''
  Phys.\ Lett.\  B {\bf 387}, 43 (1996)
  [arXiv:hep-ph/9606423].
  %%CITATION = PHLTA,B387,43;%%
  %%CITATION = PHLTA,B388,241;%%


%\cite{Svrcek:2006yi}
\bibitem{Svrcek:2006yi}
  P.~Svrcek and E.~Witten,
  ``Axions in String Theory,''
  JHEP {\bf 0606}, 051 (2006)
  [arXiv:hep-th/0605206].
  %%CITATION = JHEPA,0606,051;%%


%\cite{Komargodski:2010rb}
\bibitem{Komargodski:2010rb}
  Z.~Komargodski and N.~Seiberg,
  ``Comments on Supercurrent Multiplets, Supersymmetric Field Theories and
  Supergravity,''
  arXiv:1002.2228 [hep-th].
  %%CITATION = ARXIV:1002.2228;%%

%\cite{Copeland:1994vg}
\bibitem{Copeland:1994vg}
  E.~J.~Copeland, A.~R.~Liddle, D.~H.~Lyth, E.~D.~Stewart and D.~Wands,
  ``False Vacuum Inflation with Einstein Gravity,''
  Phys.\ Rev.\  D {\bf 49}, 6410 (1994)
  [arXiv:astro-ph/9401011].
  %%CITATION = PHRVA,D49,6410;%%


%\cite{Dimopoulos:2005ac}
%\bibitem{Dimopoulos:2005ac}
 % S.~Dimopoulos, S.~Kachru, J.~McGreevy and J.~G.~Wacker,
 % ``N-flation,''
 % JCAP {\bf 0808}, 003 (2008)
 % [arXiv:hep-th/0507205].
  %%CITATION = JCAPA,0808,003;%%

  
  %\cite{Blumenhagen:2009qh}
\bibitem{Blumenhagen:2009qh}
  R.~Blumenhagen, M.~Cvetic, S.~Kachru and T.~Weigand,
  ``{\small D}-Brane Instantons in Type {II} Orientifolds,''
  Ann.\ Rev.\ Nucl.\ Part.\ Sci.\  {\bf 59}, 269 (2009)
  [arXiv:0902.3251 [hep-th]].
  %%CITATION = ARNUA,59,269;%%

%\cite{Hosotani:1983xw}
\bibitem{Hosotani:1983xw}
  Y.~Hosotani,
  ``Dynamical Mass Generation By Compact Extra Dimensions,''
  Phys.\ Lett.\  B {\bf 126}, 309 (1983).
  %%CITATION = PHLTA,B126,309;%%

  
  %\cite{Poland:2009yb}
\bibitem{Poland:2009yb}
  D.~Poland and D.~Simmons-Duffin,
  ``Superconformal Flavor Simplified,''
  arXiv:0910.4585 [hep-ph].
  %%CITATION = ARXIV:0910.4585;%%
  
  %\cite{Kachru:2003aw}
\bibitem{KKLT}
  S.~Kachru, R.~Kallosh, A.~D.~Linde and S.~P.~Trivedi,
  ``De Sitter Vacua in String Theory,''
  Phys.\ Rev.\  D {\bf 68}, 046005 (2003)
  [arXiv:hep-th/0301240].
  %%CITATION = PHRVA,D68,046005;%%

  
  
  %\cite{McAllister:2005mq}
\bibitem{McAllister:2005mq}
  L.~McAllister,
  ``An Inflaton Mass Problem in String Inflation from Threshold Corrections  to
  Volume Stabilization,''
  JCAP {\bf 0602}, 010 (2006)
  [arXiv:hep-th/0502001].
  %%CITATION = JCAPA,0602,010;%%
  
  %\cite{Baumann:2006th}
\bibitem{BDKMMM}
  D.~Baumann, A.~Dymarsky, I.~R.~Klebanov, J.~M.~Maldacena, L.~P.~McAllister and A.~Murugan,
  ``On D3-brane Potentials in Compactifications with Fluxes and Wrapped
  D-branes,''
  JHEP {\bf 0611}, 031 (2006)
  [arXiv:hep-th/0607050].
  %%CITATION = JHEPA,0611,031;%%
  
  %\cite{Berg:2004ek}
\bibitem{BHK}
  M.~Berg, M.~Haack and B.~Kors,
  ``Loop Corrections to Volume Moduli and Inflation in String Theory,''
  Phys.\ Rev.\  D {\bf 71}, 026005 (2005)
  [arXiv:hep-th/0404087].
  %%CITATION = PHRVA,D71,026005;%%

  
  %\cite{Baumann:2010sx}
\bibitem{Baumann:2010sx}
  D.~Baumann, A.~Dymarsky, S.~Kachru, I.~R.~Klebanov and L.~McAllister,
  ``D3-brane Potentials from Fluxes in AdS/CFT,''
  arXiv:1001.5028 [hep-th]; 
  
  D.~Baumann, A.~Dymarsky, S.~Kachru, I.~R.~Klebanov and L.~McAllister,
  ``Compactification Effects in D-brane Inflation,''
  arXiv:0912.4268 [hep-th];
  
  D.~Baumann, A.~Dymarsky, S.~Kachru, I.~R.~Klebanov and L.~McAllister,
  ``Holographic Systematics of D-brane Inflation,''
  JHEP {\bf 0903}, 093 (2009)
  [arXiv:0808.2811 [hep-th]];
  
   D.~Baumann, A.~Dymarsky, I.~R.~Klebanov and L.~McAllister,
  ``Towards an Explicit Model of D-brane Inflation,''
  JCAP {\bf 0801}, 024 (2008)
  [arXiv:0706.0360 [hep-th]];
  
   D.~Baumann, A.~Dymarsky, I.~R.~Klebanov, L.~McAllister and P.~J.~Steinhardt,
  ``A Delicate Universe,''
  Phys.\ Rev.\ Lett.\  {\bf 99}, 141601 (2007)
  [arXiv:0705.3837 [hep-th]];
  
  D.~Baumann, A.~Dymarsky, I.~R.~Klebanov, J.~M.~Maldacena, L.~P.~McAllister and A.~Murugan,
  ``On D3-brane Potentials in Compactifications with Fluxes and Wrapped
  D-branes,''
  JHEP {\bf 0611}, 031 (2006)
  [arXiv:hep-th/0607050].
  %%CITATION = ARXIV:0912.4268;%%
  
 %\cite{Krause:2007jk}
\bibitem{Krause:2007jk}
  A.~Krause and E.~Pajer,
  ``Chasing Brane Inflation in String-Theory,''
  JCAP {\bf 0807}, 023 (2008)
  [arXiv:0705.4682 [hep-th]].
  %%CITATION = JCAPA,0807,023;%% 
 
 %\cite{Burgess:2006cb}
\bibitem{Burgess:2006cb}
  C.~P.~Burgess, J.~M.~Cline, K.~Dasgupta and H.~Firouzjahi,
  ``Uplifting and Inflation with D3-branes,''
  JHEP {\bf 0703}, 027 (2007)
  [arXiv:hep-th/0610320].
  %%CITATION = JHEPA,0703,027;%%
   
   %\cite{Linde:1993cn}
\bibitem{LindeHybrid}
  A.~D.~Linde,
  ``Hybrid Inflation,''
  Phys.\ Rev.\  D {\bf 49}, 748 (1994)
  [arXiv:astro-ph/9307002].
  %%CITATION = PHRVA,D49,748;%%
   
   %\cite{Baumann:2009ds}
\bibitem{BaumannTASI}
  D.~Baumann,
  ``TASI Lectures on Inflation,''
  arXiv:0907.5424 [hep-th].
  %%CITATION = ARXIV:0907.5424;%%
   
%\cite{Intriligator:1995au}
\bibitem{Intriligator:1995au}
  K.~A.~Intriligator and N.~Seiberg,
  ``Lectures on Supersymmetric Gauge Theories and Electric-Magnetic  Duality,''
  Nucl.\ Phys.\ Proc.\ Suppl.\  {\bf 45BC}, 1 (1996)
  [arXiv:hep-th/9509066].
  %%CITATION = NUPHZ,45BC,1;%%
  

%\cite{Novikov:1983uc}
\bibitem{Novikov:1983uc}
  V.~A.~Novikov, M.~A.~Shifman, A.~I.~Vainshtein and V.~I.~Zakharov,
  ``Exact Gell-Mann-Low Function Of Supersymmetric Yang-Mills Theories From
  Instanton Calculus,''
  Nucl.\ Phys.\  B {\bf 229}, 381 (1983).
  %%CITATION = NUPHA,B229,381;%%

%\cite{Intriligator:2003jj}
\bibitem{Intriligator:2003jj}
  K.~A.~Intriligator and B.~Wecht,
  ``The Exact Superconformal R-Symmetry Maximizes $a$,''
  Nucl.\ Phys.\  B {\bf 667}, 183 (2003)
  [arXiv:hep-th/0304128].
  %%CITATION = NUPHA,B667,183;%%

%\cite{'tHooft:1974bx}
\bibitem{'tHooft:1974bx}
  G.~'t Hooft and M.~J.~G.~Veltman,
  ``One Loop Divergencies in the Theory of Gravitation,''
  Annales Poincare Phys.\ Theor.\  A {\bf 20}, 69 (1974).
  %%CITATION = AHPAA,A20,69;%%
  
  %\cite{Birrell:1982ix}
\bibitem{Birrell:1982ix}
  N.~D.~Birrell and P.~C.~W.~Davies,
  ``Quantum Fields In Curved Space,''
%\href{http://www.slac.stanford.edu/spires/find/hep/www?irn=998621}{SPIRES entry}
{\it  Cambridge, Uk: Univ. Pr. ( 1982) 340p}

%\cite{Buchbinder:1992rb}
\bibitem{Buchbinder:1992rb}
  I.~L.~Buchbinder, S.~D.~Odintsov and I.~L.~Shapiro,
  ``Effective Action in Quantum Gravity,''
%\href{http://www.slac.stanford.edu/spires/find/hep/www?irn=2762668}{SPIRES entry}
{\it  Bristol, UK: IOP (1992) 413 p}

\bibitem{AdSCFT}
  D.~Baumann, A.~Dymarsky, S.~Kachru, I.~R.~Klebanov and L.~McAllister,
  ``D3-brane Potentials from Fluxes in AdS/CFT,''
  arXiv:1001.5028 [hep-th].

%\cite{Kawasaki:2000yn}
\bibitem{Kawasaki:2000yn}
  M.~Kawasaki, M.~Yamaguchi and T.~Yanagida,
  ``Natural Chaotic Inflation in Supergravity,''
  Phys.\ Rev.\ Lett.\  {\bf 85}, 3572 (2000)
  [arXiv:hep-ph/0004243].
  %%CITATION = PRLTA,85,3572;%%
  
  %\cite{Sundrum:2009ii}
\bibitem{Sundrum:2009ii}
  R.~Sundrum and C.~M.~Wells,
  ``Warped Hybrid Inflation,''
  arXiv:0909.3254 [hep-ph].
  %%CITATION = ARXIV:0909.3254;%%
  
%\cite{Blumenhagen:2005mu}
\bibitem{Blumenhagen:2005mu}
  R.~Blumenhagen, M.~Cvetic, P.~Langacker and G.~Shiu,
  ``Toward Realistic Intersecting D-brane Models,''
  Ann.\ Rev.\ Nucl.\ Part.\ Sci.\  {\bf 55}, 71 (2005)
  [arXiv:hep-th/0502005].
  %%CITATION = ARNUA,55,71;%%

%\cite{Douglas:2006es}
\bibitem{Douglas:2006es}
  M.~R.~Douglas and S.~Kachru,
  ``Flux Compactification,''
  Rev.\ Mod.\ Phys.\  {\bf 79}, 733 (2007)
  [arXiv:hep-th/0610102].
  %%CITATION = RMPHA,79,733;%%

%\cite{Blumenhagen:2006ci}
\bibitem{Blumenhagen:2006ci}
  R.~Blumenhagen, B.~Kors, D.~Lust and S.~Stieberger,
  ``Four-dimensional String Compactifications with D-Branes, Orientifolds  and
  Fluxes,''
  Phys.\ Rept.\  {\bf 445}, 1 (2007)
  [arXiv:hep-th/0610327].
  %%CITATION = PRPLC,445,1;%%

  %\cite{Denef:2008wq}
\bibitem{Denef:2008wq}
  F.~Denef,
  ``Les Houches Lectures on Constructing String Vacua,''
  arXiv:0803.1194 [hep-th].
  %%CITATION = ARXIV:0803.1194;%%
  
\end{thebibliography}
\end{document}